\documentclass[preprints,article,accept,moreauthors,pdftex,10pt]{mdpi}
\firstpage{1}
\pubvolume{xx}
\issuenum{1}
\articlenumber{5}
\pubyear{2018}
\copyrightyear{2018}
\history{Received: date; Accepted: date; Published: date}
\usepackage{caption}
\usepackage{subcaption}
\usepackage{graphicx}
\usepackage{CJKutf8}
\usepackage{CJK}

\Title{Highly dilute gas flows over an ellipse}

\Author{Shiying Cai$^1$, Chunpei Cai$^1$ and Jun Li$^2$}

\AuthorNames{Shiying Cai, Chunpei Cai and Jun Li}
\address{%
$^{1}$ \quad Department of Mechanical Engineering-Engineering Mechanics, Michigan Technological University, Houghton, 1400 Townsend Dr., Michigan 49931, U.S.A \\
$^{2}$ \quad Center for Integrative Petroleum Research, College of Petroleum Engineering \& Geosciences,  King Fahd University of Petroleum and Minerals, Dahran 31261, Saudi Arabia}

\corres{Emails: ccai@mtu.edu, junli@kfupm.edu.sa}

\abstract{This paper presents recent investigation results on  free molecular flows over a diffusely or specularly reflective ellipse, by using the gaskinetic theory. A virtual density distribution along the a diffusely reflective surface is introduced to aid the investigations. Many local surface properties are obtained, including surface slip velocity, coefficients for pressure, friction, and heat flux.  Global coefficients for aerodynamic forces and moments, mass center-force center distances, are also obtained by integrating the local surface distributions. In the end, analytical expressions for the flowfields around a diffusely or a specularly reflective ellipse are also obtained. Special non-parameters, such as temperature and speed ratios, are explicitly embedded in these expressions. Particle simulations with the direct simulation Monte Carlo (DSMC) method are performed to validate the above results. Those expressions need computers for evaluations, however, the cost is very minor when compared with DSMC simulations. The approaches are heuristic to investigate other external collisionless flows, and the load coefficients can be considered as baseline references at the high Knudsen number limit. It is feasible to further study less rarefied gas flows over an ellipse. Swift  parameter studies based on these solutions are feasible to study their influences.}

\begin{document}
%%%%%%%%%%%%%%%%%%%%%%%%%%%%%%%%%%%%%
%\setcounter{section}{-1} %% Remove this when starting to work on the template.

\section{INTRODUCTION}
Large Knudsen number (Kn) or highly rarefied gas flows may find many applications in flows in low density environments or with a small length scale. Because an ellipse shape is more general than a cylinder, rarefied gas flows over an ellipse may more properly approximate many flows. For example, gas flows over a long and elliptical asteroid, a disc alike UFO or galaxy, air flows around fine fibers inside a face mask, ellipsoidal particles such as PM2.5 or aerosols. In fact, many of such objects' cross-sections can be approximated as an ellipse. There are many investigations on compressible gas flows over an ellipse, including surface and flowfield properties. However, most past work focuses on the continuum limit with very low Reynolds numbers. There are not many theoretical studies at the free molecular flow limit, even though the geometry is very simple.  

This paper reports recent progress on investigating free molecular flows over an ellipse, and the results include surface and flowfield properties, e.g., local and total loads, pressure, friction, heat flux, moment, and the distance from the force-center to mass-center. The adopted methods include gaskinetic theory and numerical simulations. 

Because this work involves fundamental aerodynamic loads at the free molecular flow limit, it is  natural to first  review the past work on the aerodynamic forces on objects of typical shapes. In 1924, Epstein \cite{Epstein} showed the formula for the drag over a sphere, and it includes a coefficient to be determined under different flow situations. In 1946,  Tsien \cite{tsien} proposed the term ``Superaerodynamics'' for rarefied gasdynamics, and reviewed free molecular drag forces over objects with basic shapes. He presented the detailed expressions for the coefficients for pressure and friction, drag and lift forces over a flat plate. Later, Ashley \cite{ashley} reported further results on the lift and drag coefficients for free molecular flows over a symmetric airfoil, a double wedge, and a circular-arc oven, also heat flux effect is discussed even though the exact formats for the heat flux were not provided due to the complexity.  In 1948, Heineman \cite{Heineman} reported corrections to free molecular flow drags over a sphere, plate, cone or a prolated ellipsoid, and he considered molecules experienced only one collision around the object.  In 1951, Stalder and Zurick \cite{Stalder} reported  the lift and dag force coefficients for free molecular flows over a flat plate, cylinder, sphere and cone. In 1958, Gustafson \cite{Gustafson} developed the drag and moment coefficients for conic, cylindrical and spherical objects, by using the hypersonic Newtonian assumption which assumes the molecules lose their normal direction velocity and slide along the surface tangent direction. In his work, friction forces are neglected and there is no heat flux result. The collisionless Newtonian flow assumption greatly simplified the formulas with acceptable sacrifice on the accuracy. Later, in 1961, Sentman \cite{sentman} derived exact expressions of pressure, shear stress, and moment coefficients for free molecular flows over several typical and simplified satellite, including cone frustums and  spherical segments.
The heat flux coefficients for free molecular flow over a flat, plate, cone,  cylinder and a sphere were also reported \cite{Stalder_Jukoff, Oppenheim, Schaaf}.  In 1972, Wang \cite{ctwang}  reported the local loads of surface pressure, friction and heat flux over a rotating cylinder. In 2002, aerodynamic force coefficients for free molecular flows over spinning objects were reported \cite{Storch}. In 2014, gaskinetic investigations \cite{disc} on nearly collisionless flows over a disc and a sphere were reported.  

The exact solutions for free molecular flowfields around a plate, cylinder, a sphere are reported \cite{cai_plate, cai_jsr} around 2009. The resulting expressions are complex and have to be evaluated with a computer. Several simulations were performed and they validated the results. The adopted method was the direct simulation Monte Carlo (DSMC) \cite{DSMC}.  

In the literature, there are many reports on non-spherical objects, including ellipsoids, discs, and fibres. However, most of them are at the continuum and low Reynolds number flows. Most of them are also numerical simulations only. Many investigations also suggest simple relations by using curve fitting, for example, single cosine or sine function for the total drag or lift force coefficients \cite{SanjeeviS, Ouchene, Happel, Zastawny,Ouchene1, Channapan_POF}. For example, in 2019, motions of a transporting non-spherical particle in the free molecular gas flow regime are investigated \cite{Channapan_POF} by using the DSMC method, the total drag and lift force coefficients $C_D$ and $C_L$ were used, however, evidently the surface friction forces were not considered in these two coefficients. That work also presented unsteady motion relaxation procedure.  Loth \cite{Loth} provided a good review on empirical formulas for the drag and lift forces acting on non-spherical ellipsoids. 

\begin{figure}[ht]
    \begin{minipage}[l] {0.48\textwidth}
      \centering
      \includegraphics[trim=0 70 0 20, clip, height=2.4 in, width=3in]{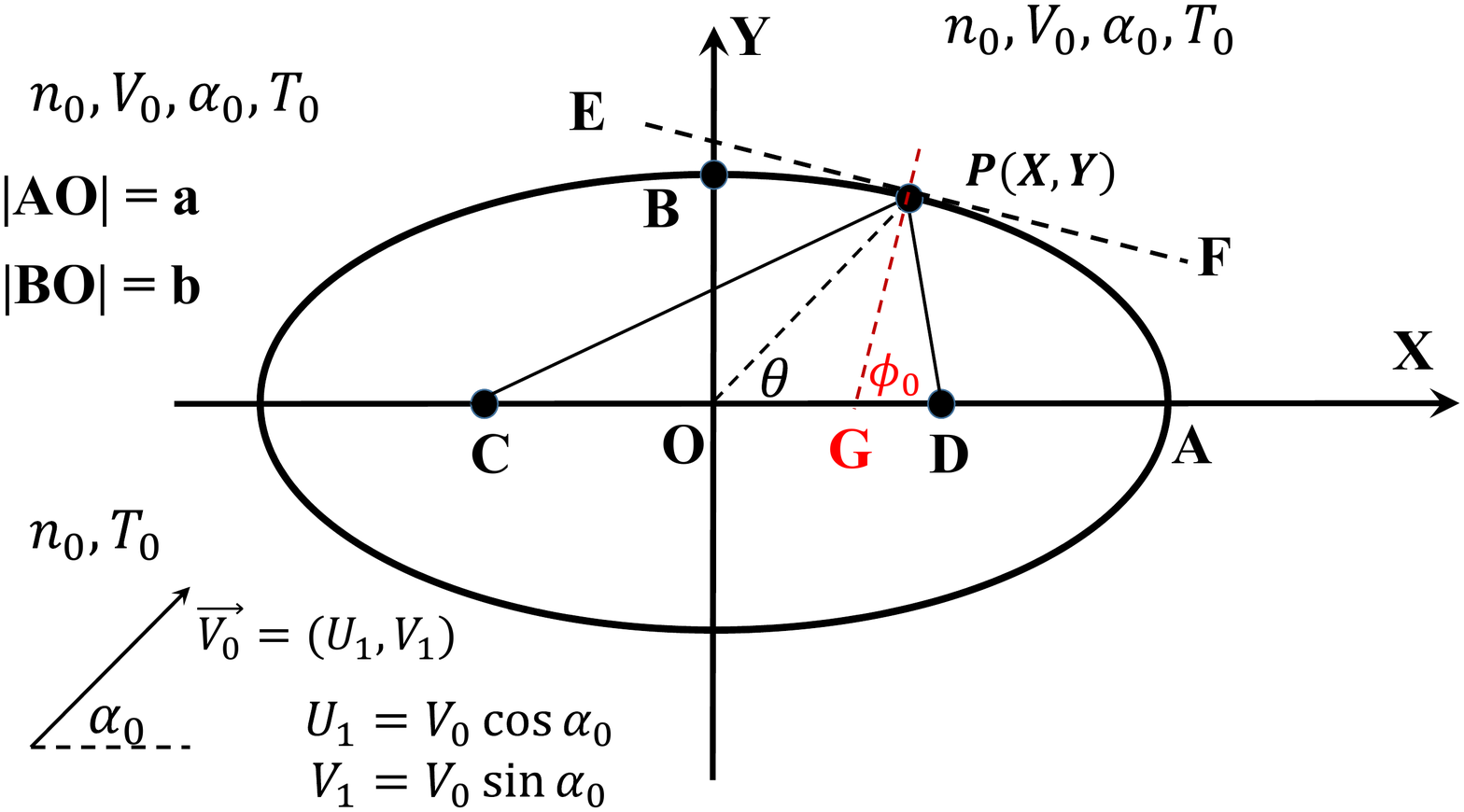}
  \end{minipage}
  \begin{minipage}[l]{0.48\textwidth}
     \centering
      \includegraphics[trim=0 70 0 20, clip, height=2.4 in, width=3in]{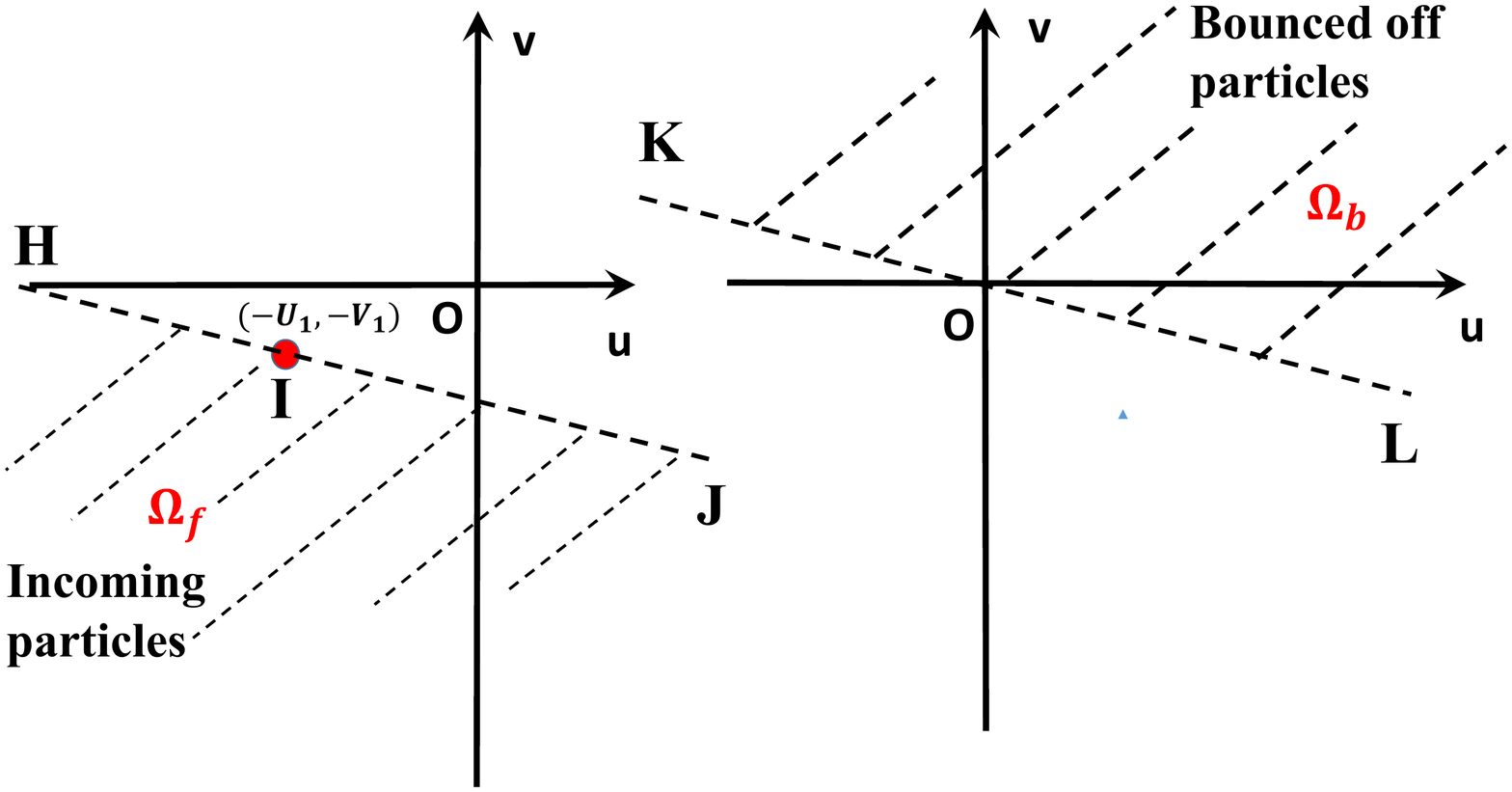}
  \end{minipage}
\caption{Problem sketch (left); and the velocity phases (right) for Point $P(X,Y)$ at a diffusely reflective surface. }
\label{Fig:problem_phase}
\end{figure}

The problem can be explained as follows. Free molecular gas flows over an  ellipse with a semi-major axis length  ``$a$'' and a semi-minor axis length ``$b$'', along the X- and Y-axis correspondingly. The surface temperature is $T_w$, and the surface can be fully diffusely or fully specularly reflective. Here, the former means when a particle hits the surface, it bounces off with a thermal velocity of uniform probabilities within a specific solid angle. The latter means that particle bounces off with the normal velocity component reversed but the tangent component unchanged. Realistic surface reflections are generally between these two limits. At the far field, the incoming flow has an Angle of Attack (AoA) $\alpha_0$, a number density  $n_0$, a velocity $V_0$, and a temperature $T_0$. For simplicity, a temperature ratio is defined as:  $\epsilon =T_w/T_0$.  The left side of Fig. \ref{Fig:problem_phase} sketches the problem. The goals of this work include: 1). develop the analytical solutions for the local surface loads; 2). discuss parameter effect on the total loads on the ellipse body; and 3). develop the exact solutions for the flowfield properties.

%[trim=0 90 0 80, clip, width=5in, height=2.3in]
%\begin{figure}[ht]
%\centering
%\includegraphics[width=6.0in]{./Figures/velocity_phase2.jpg}
%\caption{Velocity phase II.}
%\label{fig_velocity_phase2}
%\end{figure}

%\begin{figure}[ht]
%\centering
%\includegraphics[width=6.0in]{./Figures/diamend_shape.jpg}
%\caption{Velocity phase I.}
%\label{fig_velocity_phase1}
%\end{figure}

The rest of this paper is organized as follows. Section  \ref{sec:local} presents investigations on local surface properties; Section \ref{sec:global} discusses the global loads on the ellipse by using these local surface properties; Section \ref{sec:field} focuses on the flowfield property solutions; and in the end, Section \ref{sec:conclusion} summarizes this work with a few of conclusions. 

\section{LOCAL SURFACE PROPERTIES}
\label{sec:local}
Expressions for the local surface property distributions are crucial because practical concerned properties can be further integrated out from them, e.g., the total lift and drag forces, heat flux, and moments. In addition, the analytical expressions for the flowfield properties also need information about surface properties. As such, these surface properties are investigated first.

As shown in Fig. \ref{Fig:problem_phase}, there is a point $P(X,Y)$ at the ellipse surface, and it forms an angle $ POA=\theta =  \mbox{atan2}(Y/X)$. Here ``$\mbox{atan2}()$'' is a function extended from``$\mbox{atan}()$'' with a wider value range $(0, 2\pi)$. It can be shown that, the local normal direction at Point $P(X,Y)$, $\phi_0$, can be computed as \cite{ellipse}:
\begin{equation}
    \phi_0 (\theta) =  \mbox{atan2} \left( \frac{a^2 \sin\theta}{b^2 \cos \theta}  \right).
\end{equation}
%$Then $k_1 =\frac{a\sin \theta}{b \cos \theta}$. So we can compute the local normal angle  $\beta= atan(k_1)$ 

%If we consider infinitely long, then we only 
%need to add the w component with  $(-W_0, \infty)$ for the back bottom face, for the front face, we shall consider $(-\infty, W_0)$ case, for the elliptical side surface, the probably can be any w- value, think about the  $W_0 =0$ can, i.e, bulk surface along the Z-direction. 

%\begin{figure}[ht]
%\centering
%\includegraphics[width=6.0in]{./Figures/diamand_pic.jpg}
%\caption{Velocity phase I.}
%\label{fig_diamend_zona}
%\end{figure}
%
%\begin{figure}[ht]
%\centering
%\includegraphics[width=6.0in]{./Figures/diamend_phase.jpg}
%\caption{Velocity phase I.}
%\label{fig_diamend_phase}
%\end{figure}
%%%%%%%%%%%%%%%%
%Assuming the free stream bulk velocity $(U_0, V_0)$ which defines the absolute external flow and AoA. The free-stream temperature is $T_0$.  The surface temperature is assumed to be $T_w$.

\subsection{Diffusely reflective surface}
Figure \ref{Fig:problem_phase} shows the velocity phases for Point $P(X,Y)$ at a diffusely reflective ellipse surface. The velocity phases include two regions: one is $\Omega_b$ which is a half plane, representing those particles bouncing off from Point $P(X,Y)$, and $\Omega_f$ represents the incoming free stream molecules moving toward Point $P(X,Y)$. The angle $\angle PGD= \phi_0 $ is the local normal direction at Point $P(X,Y)$. In this figure, three lines are parallel,  $ EF \parallel HJ  \parallel KL$. Line $KL$ passes the coordinate origin, and  $HJ$ passes a special Point $I$ with coordinates $(-U_1, -V_1)$ with $U_1=V_0 \cos \alpha_0$ and $V_1= V_0 \sin \alpha_0$. More details about how to construct these velocity phases can be found in two previous publications \cite{cai_jsr, impingement}.

To aid analyzing free molecular flows over a diffusely reflective ellipse, first a ``virtual'' Maxwellian velocity distribution function (VDF) is introduced for area $\Omega_b$:
\begin{equation}
     f_w(u,v) = n_w (\theta) \sqrt{ \frac{\beta}{\pi}  } e^{-\beta_w (u^2+v^2)},
\label{eqn:vdfw}
\end{equation}
where $\beta_w = 1/(2RT_w)$, and  $n_w(\theta)$ is a ``virtual'' number density which can be determined by using the non-penetration boundary conditions at the Point $P(X,Y)$. This VDF and $\Omega_b$ can sufficiently describe those particles bouncing off from Point $P(X,Y)$ at the diffusely reflective ellipse surface. 
\begin{figure}[ht]
    \begin{minipage}[l] {0.48\textwidth}
      \centering
      \includegraphics[width=3.6in]{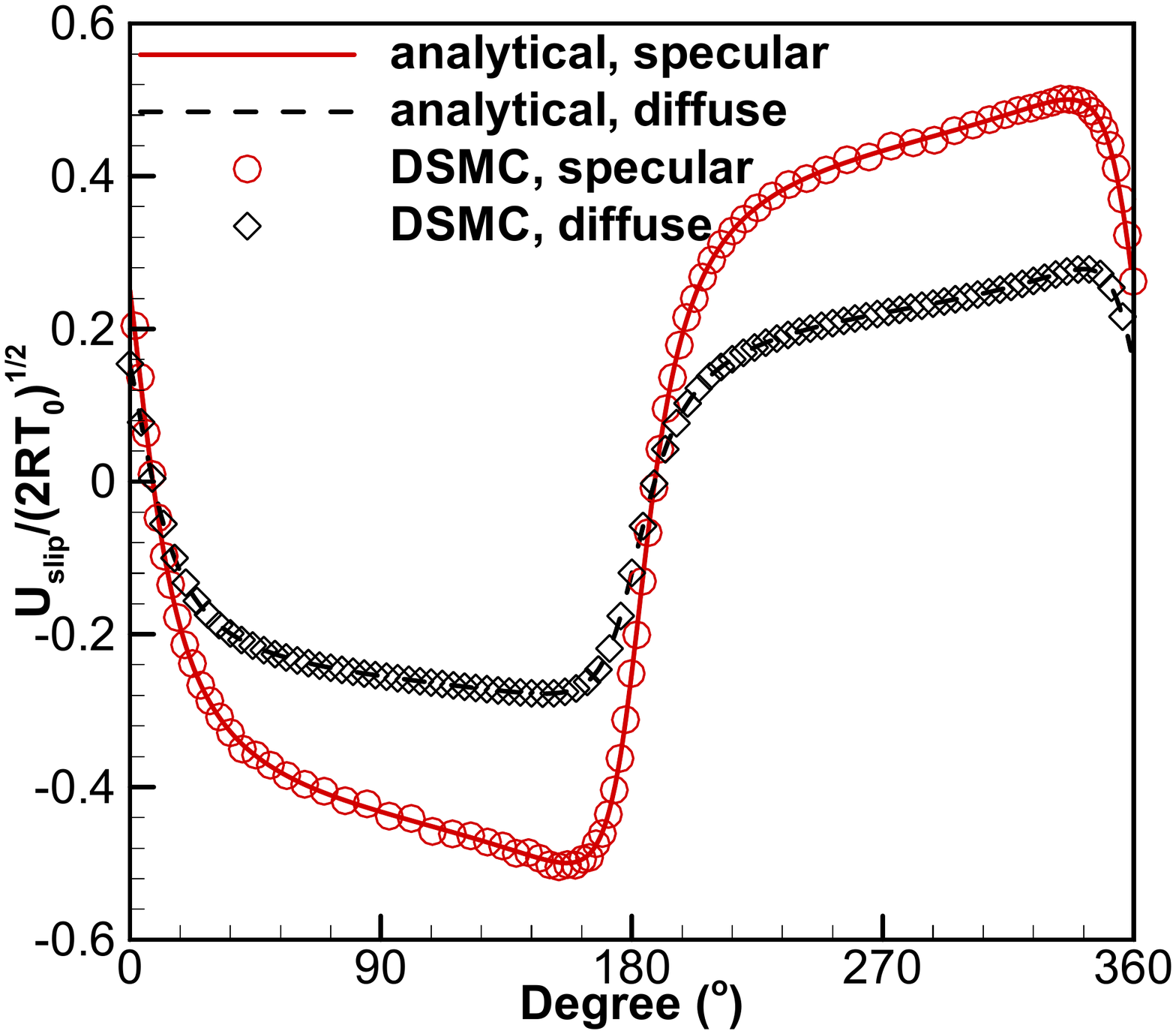}
  \end{minipage}
  \begin{minipage}[l]{0.48\textwidth}
     \centering
      \includegraphics[width=3.6in]{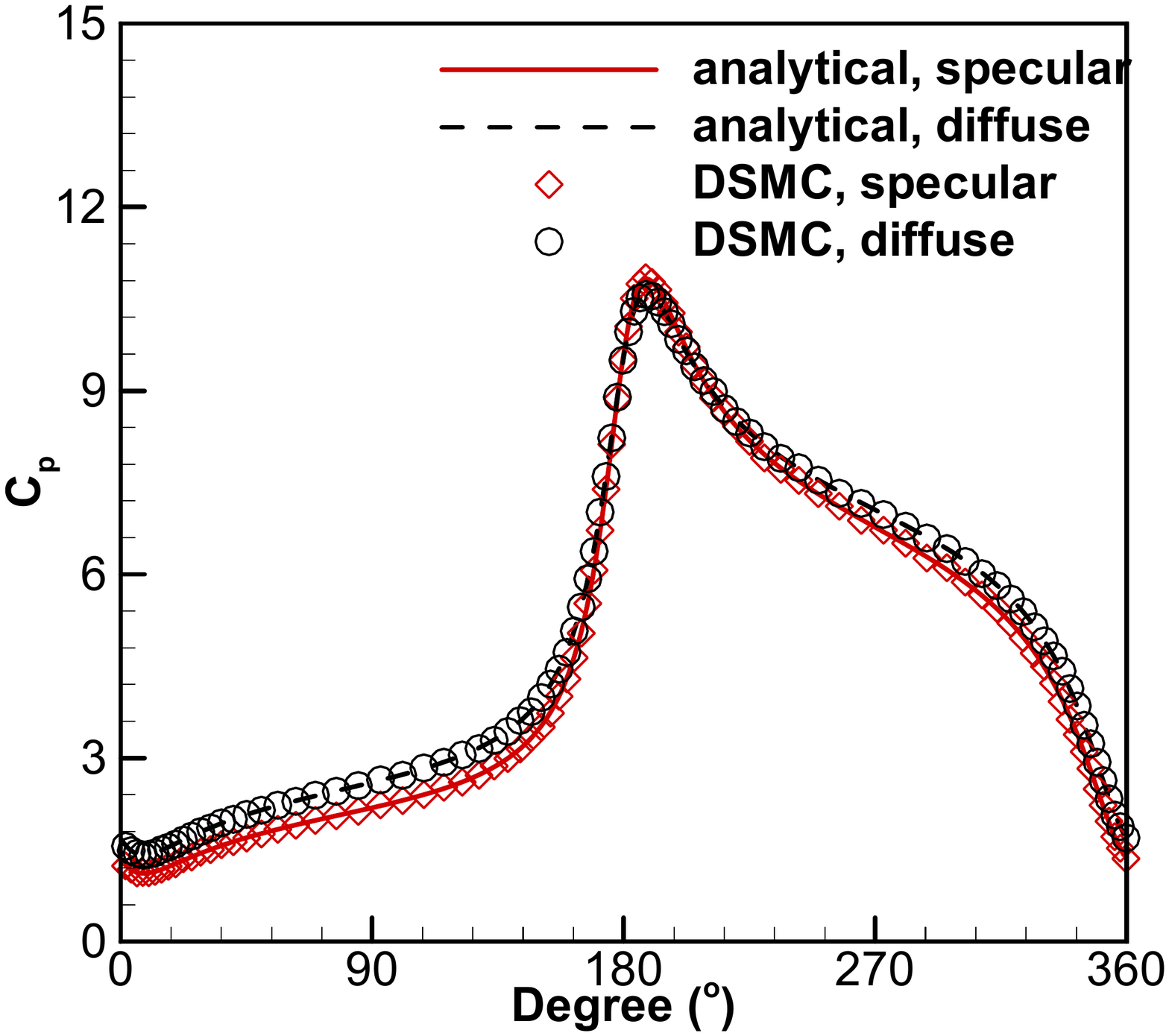}
  \end{minipage}
\caption{Surface slip velocities, $u_{slip}(\theta)/{\sqrt{2RT_0} }$ (left), and surface pressure coefficient $p_w/(\rho_0 V^2_0/2)$ (right), along a diffusely or specularly reflective ellipse. Symbols: DSMC; Lines: analytical. $a/b=2$, $\alpha_0=30^\circ$, $S_0=0.5$, $a/b=2$, and $\epsilon =1.5$.}
\label{Fig:slipu_Cp}
\end{figure}
\begin{figure}[ht]
    \begin{minipage}[l] {0.48\textwidth}
      \centering
      \includegraphics[width=3.6in]{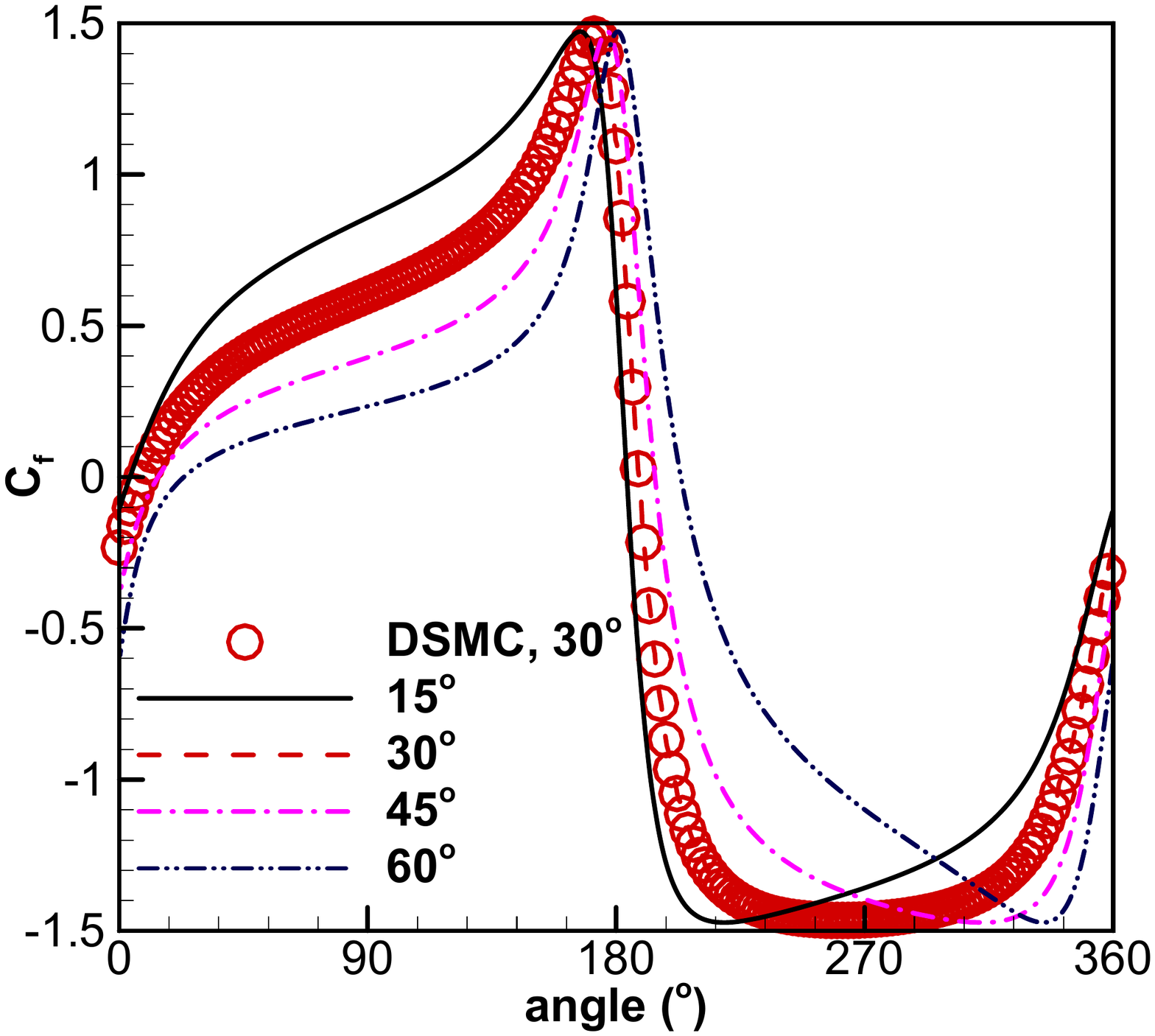}
  \end{minipage}
  \begin{minipage}[l]{0.48\textwidth}
     \centering
      \includegraphics[width=3.6in]{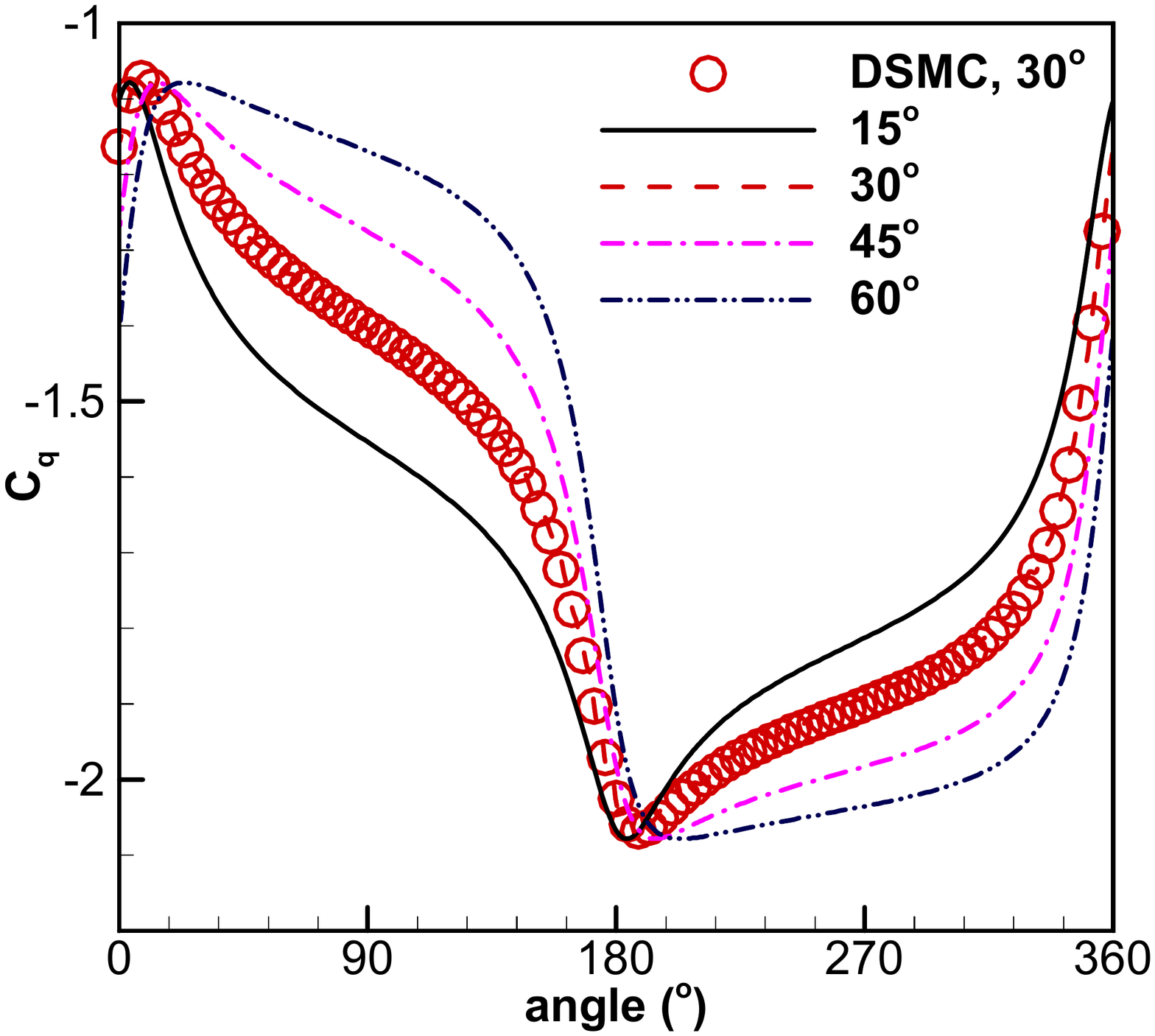}
  \end{minipage}
\caption{Surface friction coefficient, $f_w/(n_0m V_0^2/2)$ (left), and surface heat flux coefficient, $q_w/(n_0 m V_0^3/2)$ (right), along a diffusely reflective ellipse. Symbol: DSMC; Lines: analytical. $a/b=2$, $\alpha_0 =15^\circ$, $30^\circ$, $45^\circ$, $60^\circ$, $S_0=0.5$, and $\epsilon =1.5$.}
\label{Fig:Cf_Cq}
\end{figure}

By using the gaskinetic theory (e.g.,\cite{Kogan}), Eqn. \ref{eqn:vdfw}, and $\Omega_b$, the out-coming flux formed by those bounced off particles from $P(X,Y)$ can be integrated out as: 
\begin{equation}
    n_w (\theta) \int_0^{\infty} u \sqrt{  \frac{\beta_w}{\pi} } e^{-\beta_w u^2} du = \frac{n_w (\theta)}{2 \sqrt{\beta_w \pi}}.
\label{eqn:fluxwall}
\end{equation}

The free stream molecules can be described by another Maxwellian VDF characterized by $n_0$, $T_0$, $\alpha_0$, $V_0$, the  velocity phase is $\Omega_f$, and a half plane which does not pass the coordinate origin. The incoming flux from the free stream to Point $P(X,Y)$ at the diffusely reflective ellipse surface can be computed by using this Maxwellian VDF, $\Omega_f$, and the gaskinetic theory. The result is:
\begin{equation}
  n_0 \frac{\beta_0}{\pi} \int_{\Omega_o} [(u+U_1) \cos \phi_0 +(v+V_1) \sin \phi_0 ] e^{-\beta_0 (u^2+v^2)} dudv =
    n_0    \frac{ \beta_0}{\pi} e^{-S_0^2} \int\limits_{\phi_0 +\pi/2}^{\phi_0 +3\pi/2} \cos(\eta -\phi_0) e^{K^2} C^* d\eta, 
\label{eqn:fluxin}
\end{equation}
where $K\equiv S_0\cos(\eta-\alpha_0)$, $S_0 = V_0 /\sqrt{2RT_0}$, and $C^*(K)$ is a function and its expression is included in the appendix. 

The expression for the ``virtual'' number density $n_w(X,Y)$ for Point $P(X,Y)$ can be determined from Eqns.  
 \ref{eqn:fluxwall} and \ref{eqn:fluxin}:
\begin{equation}
     \frac{n_w(\theta)}{n_0}  = 
      2e^{-S_0^2} \sqrt{ \frac{\epsilon}{\pi}}  \int_{\phi_0 +\pi/2}^{\phi_0 +3\pi/2} \cos(\eta -\phi_0(\theta)) e^{K^2}   C^*(K) d\eta. 
\label{eqn:nw}
\end{equation}
%%%%%%%%%%%%%%%%%%%%%%%%%%%%%%%%
The normalized number density at Point $P(X,Y)$ consists of two parts, $n_w(\theta)$, and $n_{in}$. The latter is due to the free molecular free-stream, with $\Omega_f$ and the gaskinetic theory, it can be integrated out as:  
\begin{equation}
    \frac{ n_{in} }{ n_0} = \frac{e^{-S_0^2}}{\pi} \int_{\phi_0 +\pi/2}^{\phi_0 +3\pi/2} e^{K^2}   B^*(K) d\eta,
\end{equation}
where the expression for $B^*$ is available in the appendix as well.  

The momentum along a diffusely reflective ellipse surface is completely due to the free-stream flow, and the $n_w(X,Y)$ does not contribute to the momentum. At Point $P(X,Y)$, the slip velocity  along the surface tangent direction is: 
\begin{equation}
    \frac{ u_{slip,d}(\theta) }{ \sqrt{2RT_0}} = \frac{n_0}{n_{in}+n_w(\theta)/2} \frac{e^{-S_0^2}}{\pi} \int_{\phi_0 +\pi/2}^{\phi_0 +3\pi/2} \sin(\eta -\phi_0(\theta) ) e^{K^2}   C^*(K)  d\eta.  
\end{equation}
The slip velocity direction is defined counter-clock-wisely as $\theta$ increases. Note in the above equation, the local surface normal $\phi_0$ is a function of $\theta$, as sketched in Fig. \ref{Fig:problem_phase}. 

The coefficients for surface pressure, friction, and heat flux along a diffusely reflective ellipse can be obtained by using the gaskinetic theory, $\Omega_b$ and $\Omega_f$, and the two Maxwellian VDFs. With careful simplifications and derivations, the final expressions are:
\begin{equation}
    C_{p,d}(\theta) =\frac{2}{n_0 m V_0^2}   \int_{\Omega} C_n^2 f d C_n   = \frac{n_w (\theta) }{2n_0}\frac{\epsilon}{S_0^2} +  \frac{2 e^{-S_0^2}  }{\pi S_0^2 } \int_{\phi_0  + \pi/2}^{\phi_0 + 3 \pi/2} \cos^2(\eta-\phi_0(\theta) ) e^{K^2} D^*(K) d\eta,
\label{eqn:cpd}
\end{equation}
\begin{equation}
    C_f(\theta ) =  \frac{e^{-S_0^2}}{ \pi S_0^2}  \int_{\phi_0 +\pi/2}^{\phi_0 + 3\pi/2} \sin(2\phi_0 (\theta)- 2\eta) e^{K^2} D^*(K) d\eta,
\label{eqn:cfd}
\end{equation}
\begin{equation}
     C_q(\theta) = \frac{q_w}{n_0 m V_0^3/2} =  \frac{1}{\sqrt{\pi} \epsilon^{3/2} S_0^3} \frac{n_w(\theta)}{n_0}   +  \frac{ e^{-S_0^2} }{ \pi S_0^3}   \int_{\phi_0 + \pi/2}^{\phi_0 + 3\pi/2} \cos(\eta-\phi_0(\theta))  e^{K^2 } \bigg(  E^*(K)  +  \frac{1}{2}  C^*(K)   \bigg)  d\eta,
\label{eqn:cqd}
\end{equation}
where the expressions for integrands $C^*(K)$, $D^*(K)$ and $E^*(K)$ can be found in the appendix. As shown, the normal direction $\phi_0$ is a crucial factor embedded in the results. 

%\begin{figure}[h]
%\centering
%\includegraphics[width=4.5in]{./Figures/Cf_Cq_specular_diffuse_pdf2}
%\caption{Friction and heat flux coefficient distributions along a diffuse reflective ellipse, AoA=$30^\circ$, $S_0=0.5$, $a/b=2$, $T_0/T_w=2/3$.}
%\end{figure}
%\begin{figure}[h]
%\centering
%\includegraphics[width=4.5in]{./Figures/field_p.pdf}
%\caption{Pressure field around an ellipse, $P(x,y)/(n_0 k T_0)$, AoA=$30^\circ$, $S_0=0.5$, $a/b=2$, $T_0/T_w=2/3$, $a/b=2.$  Top: diffuse, bottom: specular; analytical: dashed; DSMC: solid.}
%\label{fig:Txy_diffuse}
%\end{figure}

%%%%%%%%%%%%%%%%%%%%%%%%end of diffuse reflective ellipse 
\subsection{Specularly reflective surface properties} 
The contribution to the number density at Point $P(X,Y$ of a specularly reflective surface consists of two parts as well, one is due to the incoming free molecular free stream, $n_{in,s}$, and the other is from those reflected particles which is exactly the same as $n_{in,s}$:    
\begin{equation}
    \frac{ n_{in, s} (\theta) }{n_0} = \frac{e^{-S_0^2}}{\pi} \int_{\phi_0 +\pi/2}^{\phi_0 +3\pi/2} e^{K^2 }  B^*(K)  d\eta. 
\end{equation}
%%%%%%%%%%%%%%%%%%%%%%%%%%%%%%%
The velocity slips and pressure coefficients along a specularly reflective ellipse surface are:
\begin{equation}
    \frac{u_{slip,s}(\theta)}{ \sqrt{2RT_0}} =  \frac{n_0}{ n_{in,s} (\theta) } \frac{e^{-S_0^2}}{\pi}       \int_{\phi_0  + \pi/2}^{\phi_0 + 3 \pi/2}  \sin( \eta - \phi_0(\theta) ) e^{K^2 }  C^*(K) d\eta,
\end{equation}
\begin{equation}
    C_{p,s}(\theta)  = \frac{p_{w,s} (\theta) }{n_0 m V_0^2/2}  =  \frac{4  e^{-S_0^2}  }{\pi S_0^2 } \int_{\phi_0  + \pi/2}^{\phi_0 + 3 \pi/2}  \cos^2( \eta-\phi_0 (\theta) ) e^{K^2 } D^*(K) d\eta.
\label{eqn:cps}
\end{equation}
According to the gaskinetic theory, the friction and heat flux along a specularly reflective surface is zero. 

To validate the above expressions, two  DSMC  simulations are performed. The simulation package \cite{GRASP} is well tested.  The simulation parameters include  $a/b=2$, $\alpha_0 = 30^\circ$, and $\epsilon =1.5$. The ellipse surface is divided into 360 degrees and 360 small segments, which can aid the surface property computations. 

Figure \ref{Fig:slipu_Cp} shows the surface slip velocities (left) and pressure coefficients (right) along a diffusely or specularly reflective surface. There are two regions with relatively larger slip velocities, where the slip velocities along a specularly reflective surface are relatively larger. For the surface pressure distributions, the peak values happen around  $\theta=210^\circ$ which is the impingement angle, with an incoming flow  $\alpha_0 = 30^\circ$.   

Figure \ref{Fig:Cf_Cq} shows surface friction and heat flux coefficients along a diffuse reflective surface. For the $\alpha_0 = 30^\circ$ case, the DSMC simulation results and the analytical solutions essentially are identical. Also included in this figure are several other curves with different $\alpha_0$, to illustrate the variation trends for $C_f$ and $C_q$ along a diffusely reflective surface. As shown, with different $\alpha_0$, the maximum and minimum $C_f$ and $C_q$ values remain constant, but their positions on the ellipse shift correspondingly. This is because the ellipse surface can be considered as many small segments or panels, with different $\alpha_0$, the most sensitive impingement point monotonically shifts to the neighboring segment with a different inclination angle. This may generally correct for free molecular external flows over any non-convex objects which include but not limited to ellipses.  

\section{GLOBAL LOADS OVER AN ELLIPSE}
\label{sec:global}
The exact expressions for surface local properties are validated in the previous section.  These surface properties can lead to global properties by using numerical integration, and there is no need of simulations anymore.  This is one of one great advantage of this work.

There are several important coefficients,  including the total lift ($C_L(\alpha_0)$) and drag ($C_D(\alpha_0)$) forces, moment ($C_M(\alpha_0)$), and the distance from the force-center to the mass-center ($S_{cc}(\alpha_0)$):
\begin{equation}
       C_D(\alpha_0) = \frac{1}{2a}    \int_0^{2\pi}  \left(
         C_f(\theta) \sin \phi_0(\theta)  +C_p(\theta) \cos \phi_0(\theta)  
         \right)  L(\theta) d\theta, 
\label{eqn:CD}
\end{equation}
\begin{equation}
       C_L(\alpha_0) = \frac{1}{2a} \int_0^{2\pi}   \left( 
         C_f (\theta) \cos \phi_0(\theta) -C_p(\theta) \sin \phi_0(\theta) 
       \right)  L(\theta) d\theta,
\label{eqn:CL}
\end{equation}
\begin{equation}
       C_M(\alpha_0) =  \frac{1}{4a^2}   \int_0^{2\pi}  \left( C_f(\theta)  \cos(\theta -  \phi_0)  + C_p(\theta) \sin(\theta - \phi_0)   \right)  L^2(\theta) d\theta.
\label{eqn:CM}
\end{equation}
The normalized distance from the force-center to the mass-center is defined as:
\begin{equation}
   S_{cc}(\alpha_0)= C_M(\alpha_0) /{\sqrt{C_D^2(\alpha_0) +C_L^2(\alpha_0) }}
\label{eqn:CCD}
\end{equation}
where $2a$ is the longest length chosen for normalization, and $L(\theta) = ab/\sqrt{ a^2 \sin^2 \theta + b^2 \cos^2 \theta} = \sqrt{X^2 + Y^2}$ is the distance from Point $P(X,Y)$ on the ellipse surface to the ellipse center. The drag and lift forces are defined as along the X- and Y+ directions; and moment is defined positive along the counterclockwise direction. 

As shown, the expressions for the local and global load distributions include several ratios: $\alpha_0$, $S_0$,  $a/b$, and $T_w/T_0$ which is for the diffusely reflective ellipse only and relatively not important. Hence, we can evaluate the parameter's effect from $a/b$ and $S_0$. 

\begin{figure}[ht]
    \begin{minipage}[l] {0.48\textwidth}
      \centering
      \includegraphics[width=3.6in]{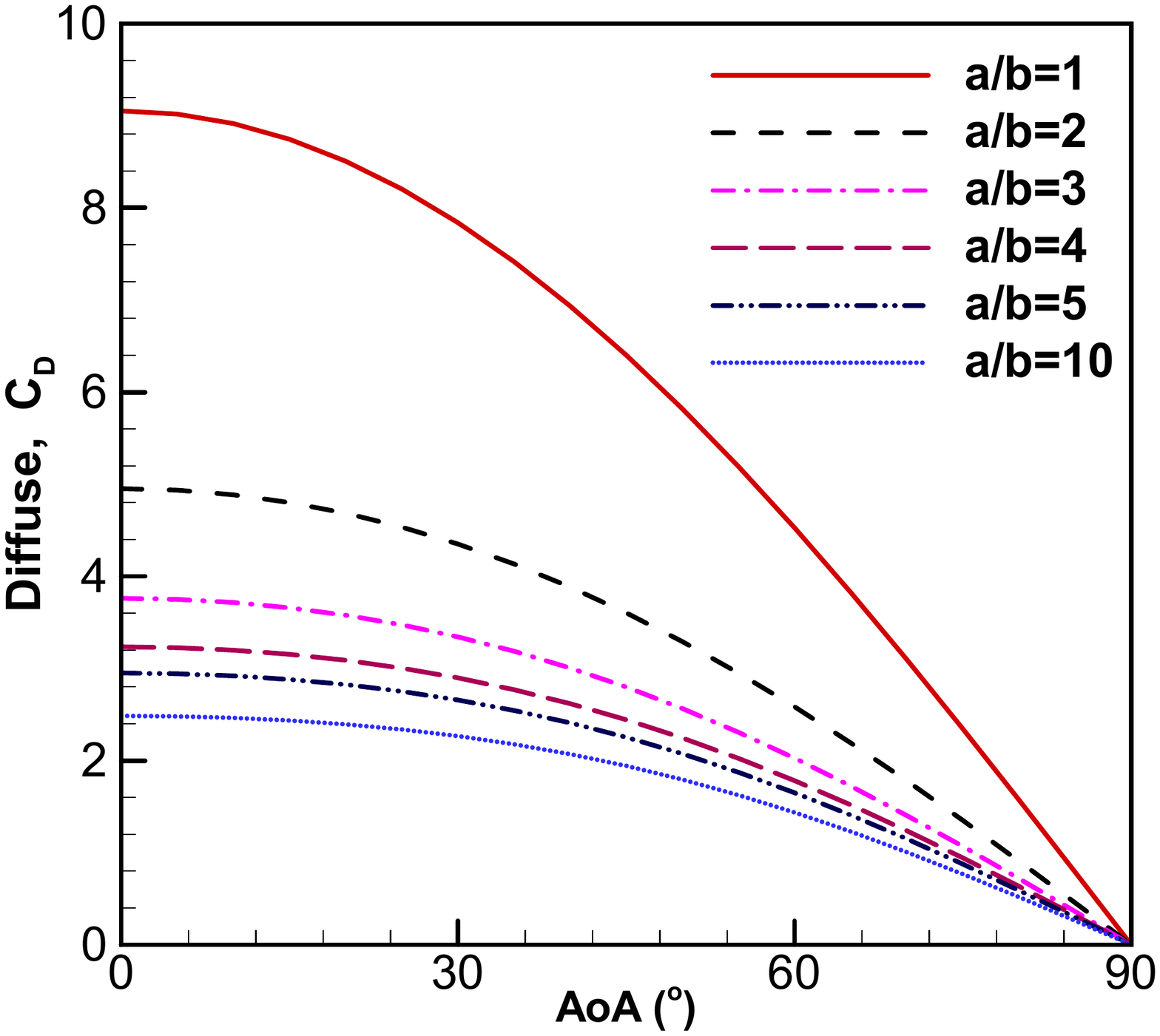}
  \end{minipage}
  \begin{minipage}[l]{0.48\textwidth}
     \centering
      \includegraphics[width=3.6in]{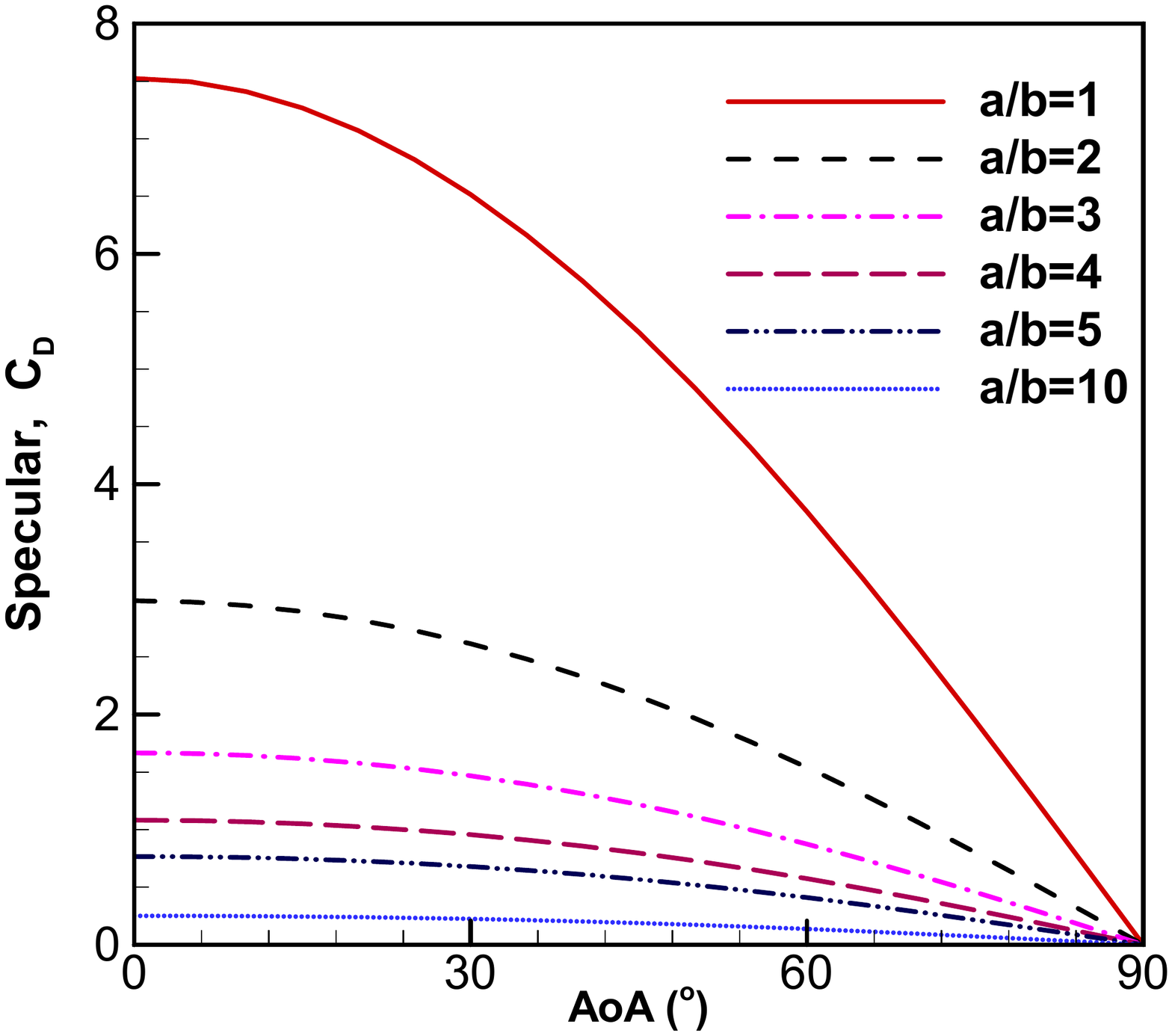}
  \end{minipage}
\caption{Aspect ratio  effect on $C_D$, diffuse (left) and specularly (right) reflective surfaces. $S_0=0.5$, and $\epsilon =1.5$.}
\label{Fig:CD_alpha}
\end{figure}

\begin{figure}[ht]
    \begin{minipage}[l] {0.48\textwidth}
      \centering
      \includegraphics[width=3.6in]{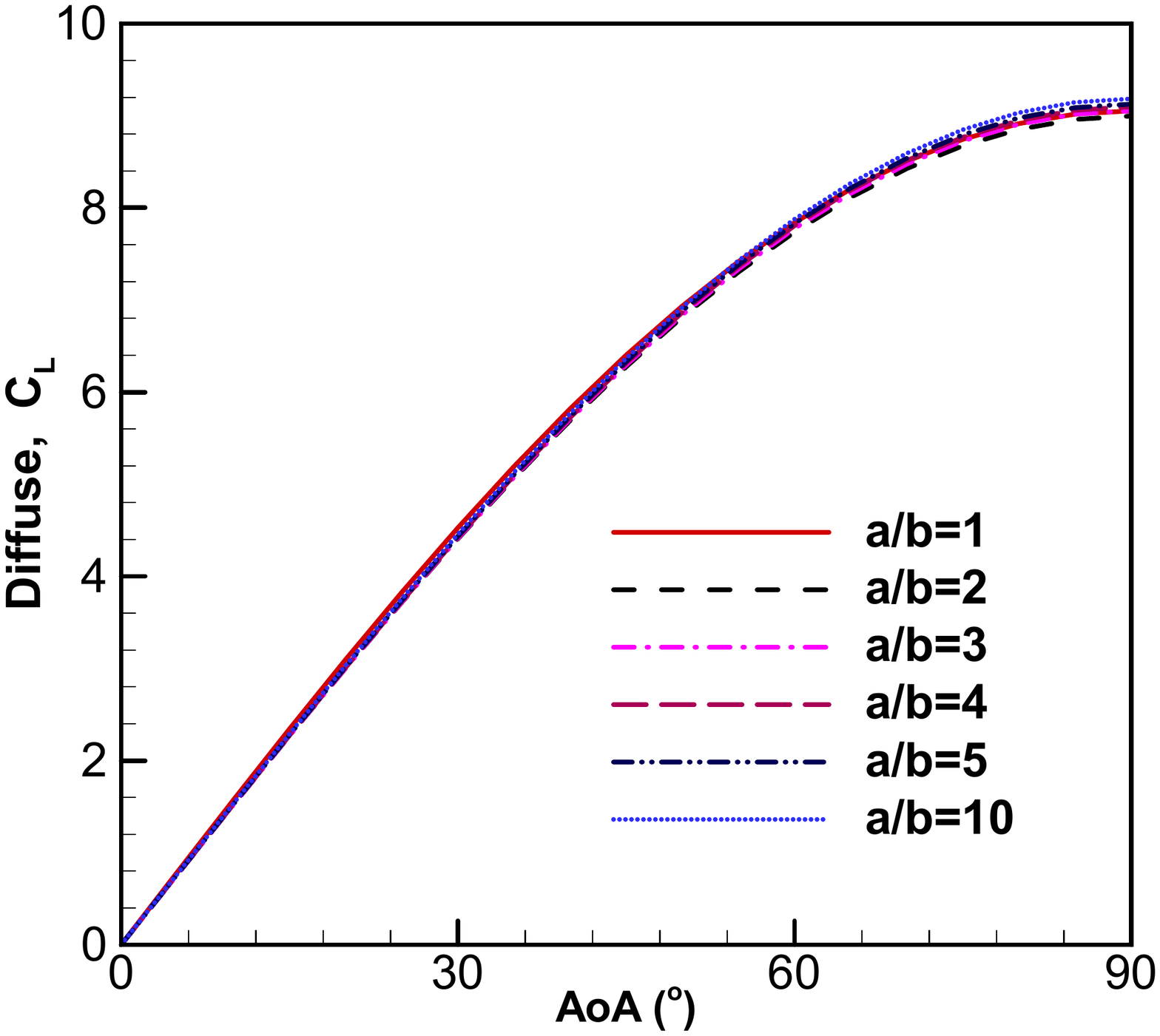}
  \end{minipage}
  \begin{minipage}[l]{0.48\textwidth}
     \centering
      \includegraphics[width=3.6in]{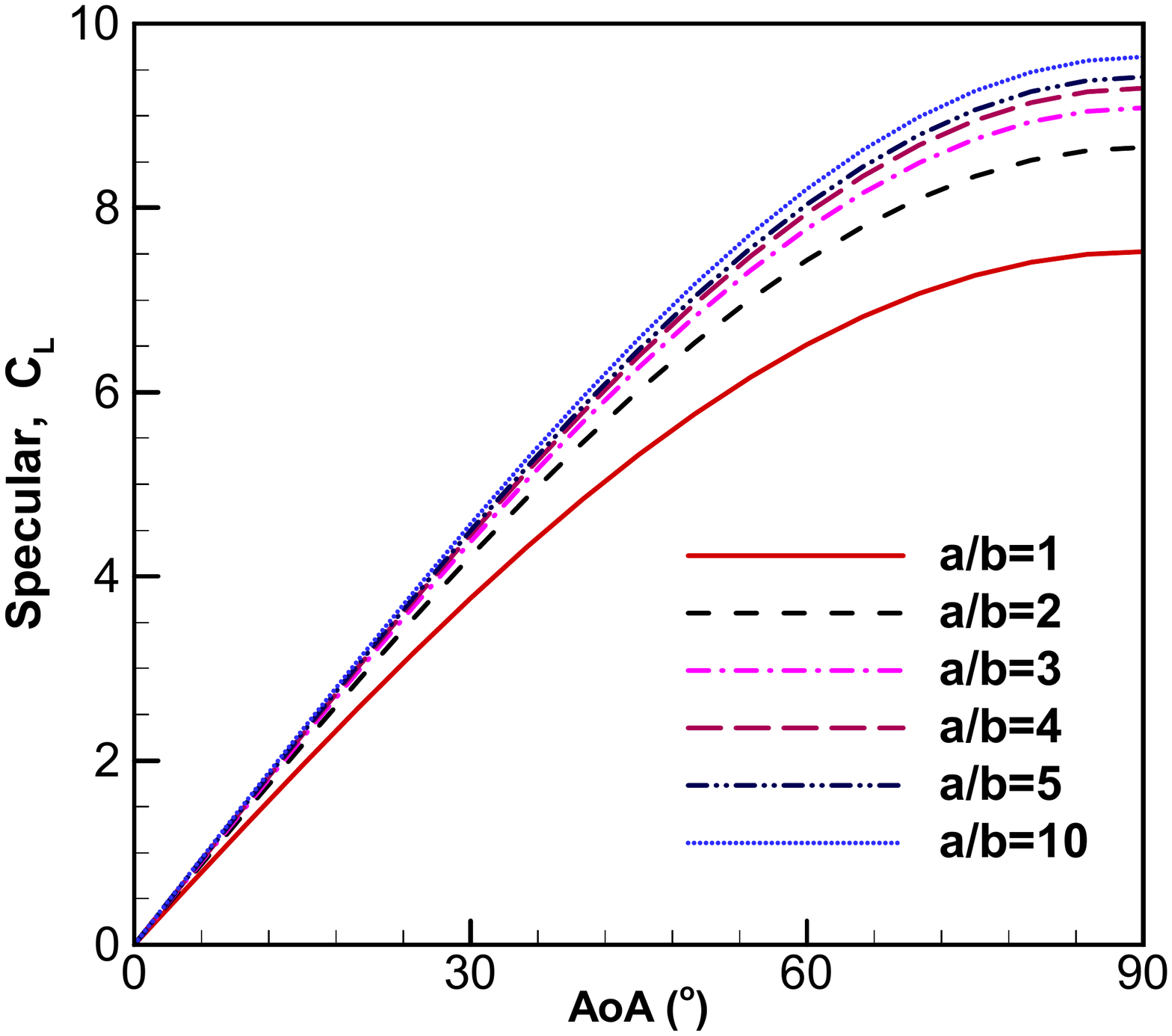}
  \end{minipage}
\caption{Aspect ratio effect on $C_L$, diffuse (left) or specularly (right) reflective surfaces. $S_0=0.5$, and $\epsilon =1.5$.}
\label{Fig:CL_alpha}
\end{figure}

\begin{figure}[ht]
    \begin{minipage}[l] {0.48\textwidth}
      \centering
      \includegraphics[width=3.6in]{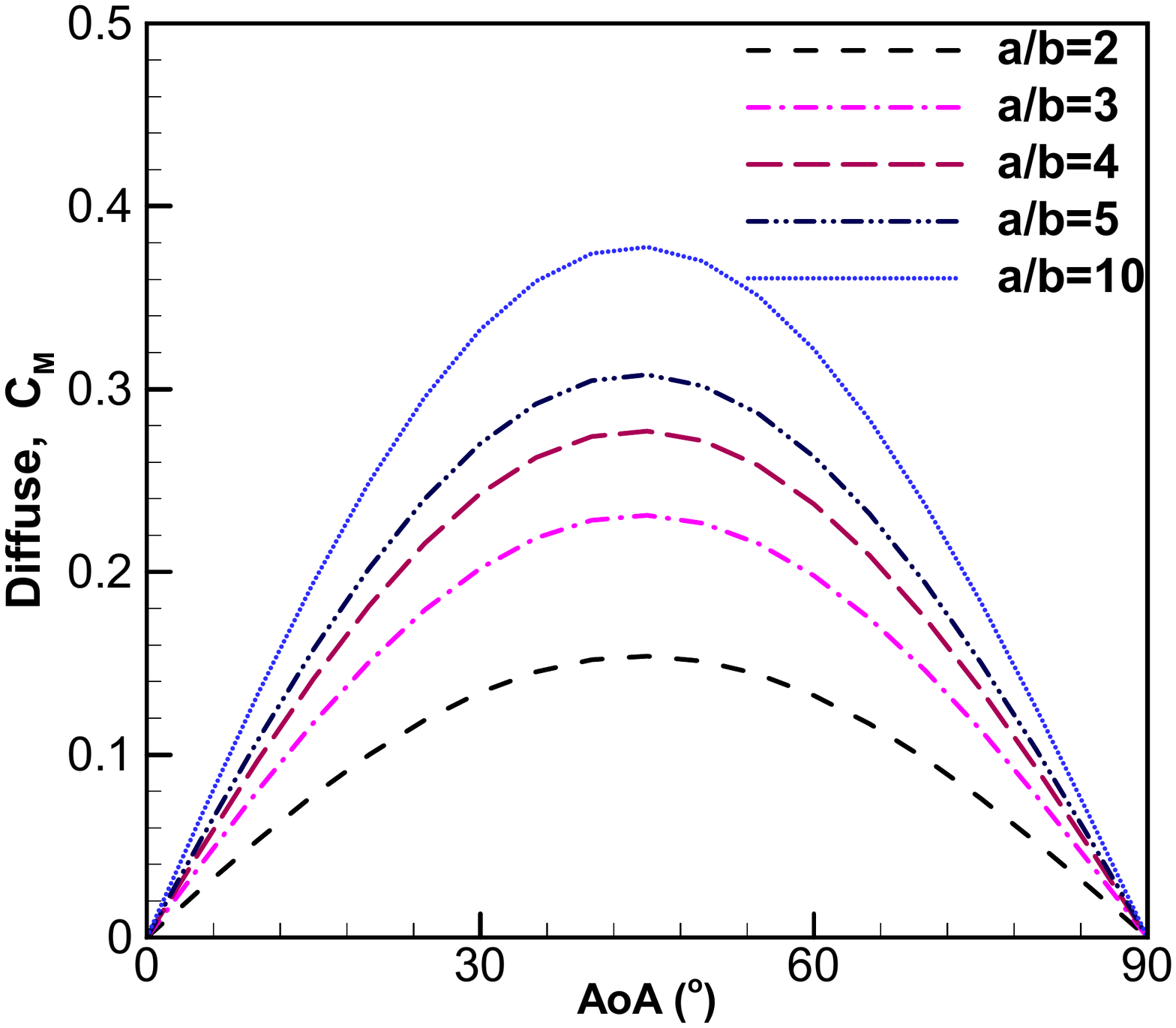}
  \end{minipage}
  \begin{minipage}[l]{0.48\textwidth}
     \centering
      \includegraphics[width=3.6in]{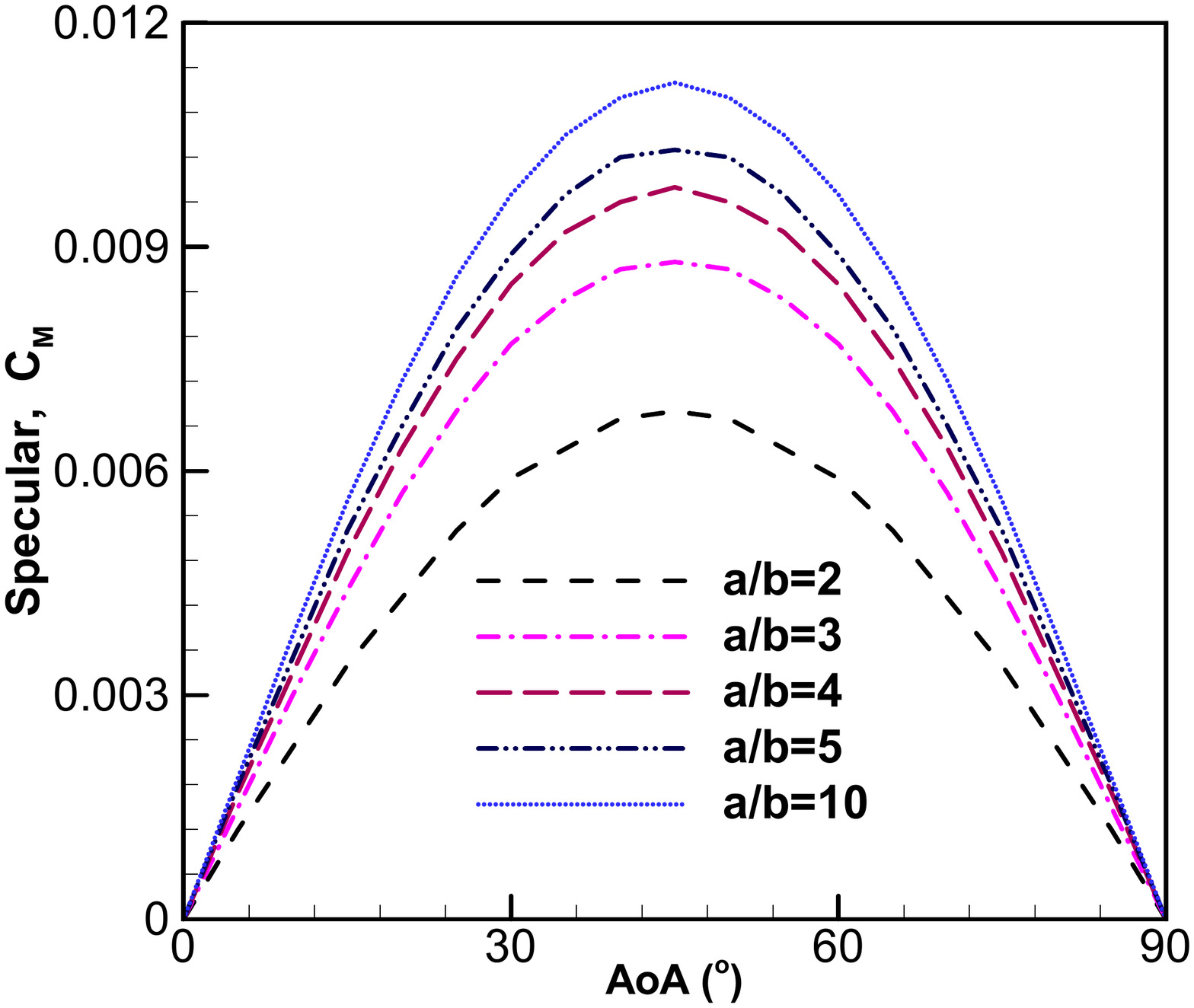}
  \end{minipage}
\caption{Aspect ratio effect on $C_M$, diffuse (left) or specularly (right) reflective surface.  $S_0=0.5$, and $\epsilon =1.5$.}
\label{Fig:CM_alpha}
\end{figure}

\begin{figure}[ht]
    \begin{minipage}[l] {0.48\textwidth}
      \centering
      \includegraphics[width=3.6in]{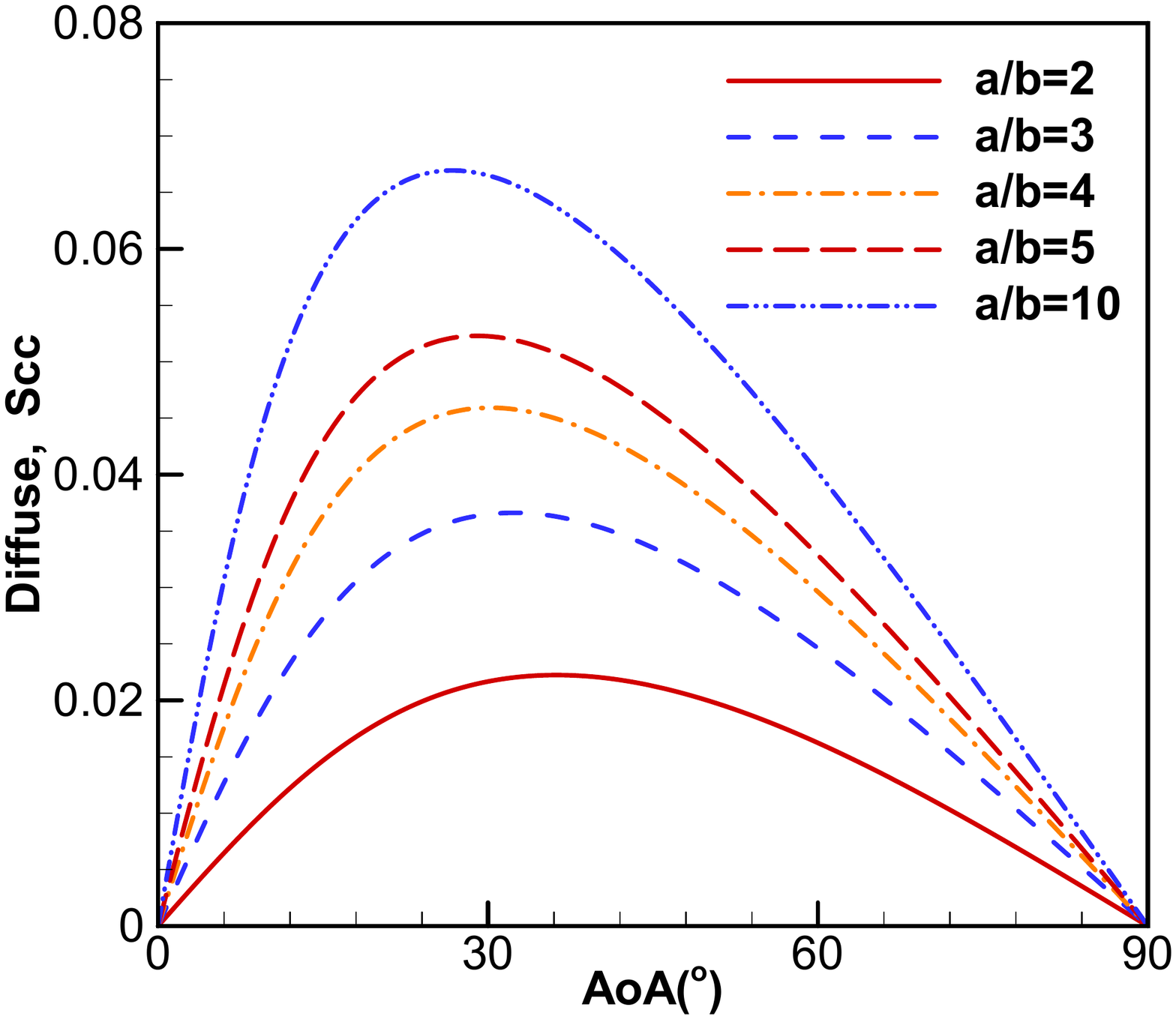}
  \end{minipage}
  \begin{minipage}[l]{0.48\textwidth}
     \centering
      \includegraphics[width=3.6in]{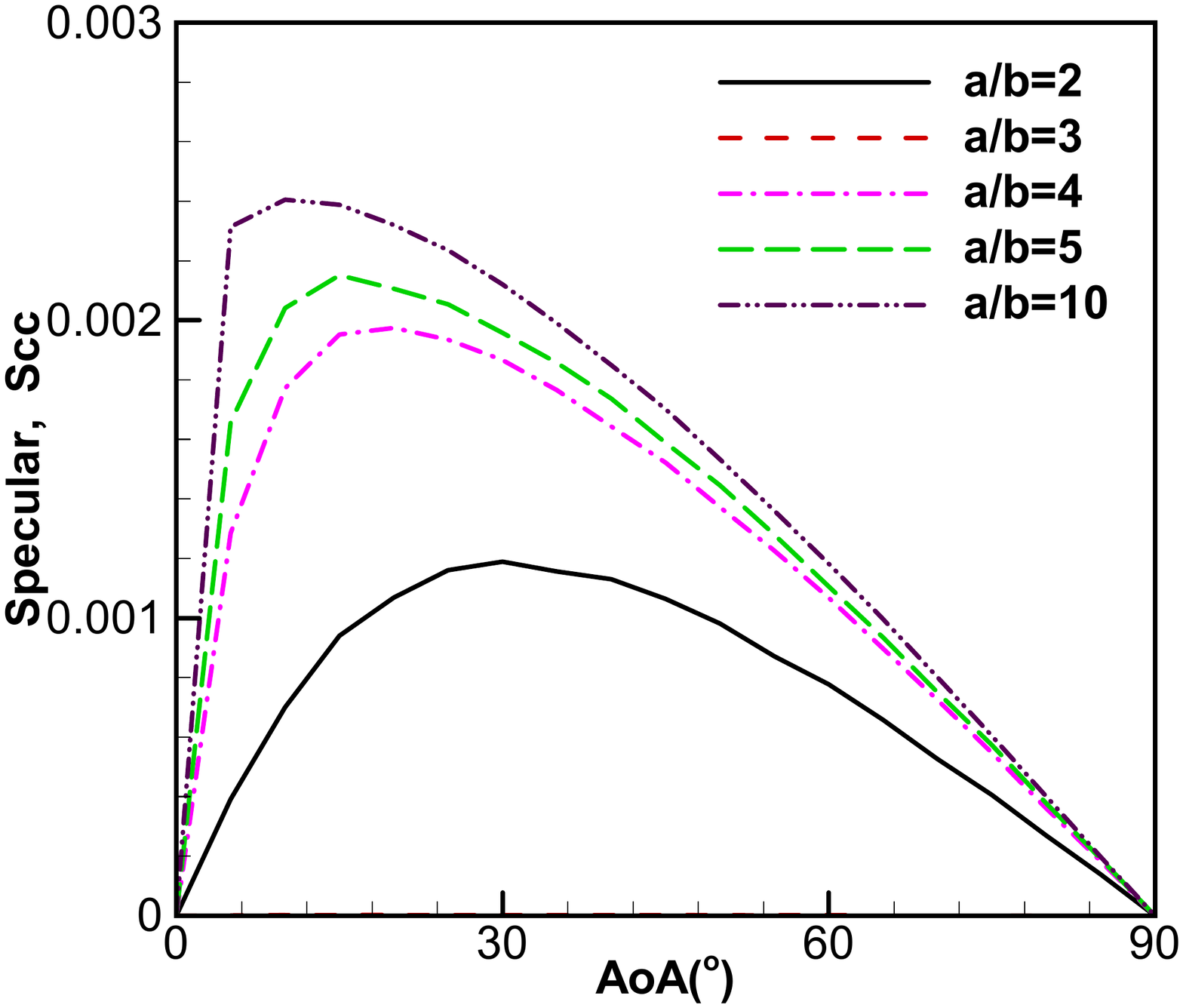}
  \end{minipage}
\caption{Aspect ratio (left) effect on $S_{cc}$, diffuse (left) and specular (right) ellipse, $S_0=0.5$ and $\epsilon =1.5.$}
\label{Fig:scc_abratio_diffuse_specular}
\end{figure}

First, the $a/b$ ratio effect on these four properties are shown as Figs.\ref{Fig:CD_alpha},  \ref{Fig:CL_alpha}, \ref{Fig:CM_alpha} and \ref{Fig:scc_abratio_diffuse_specular}. As the aspect ratio $a/b$ increases, 1). The ellipse becomes flatter, more like a rounded flat plate; 2). The ellipse area decrease;  3). The arm $L(\theta)$ becomes smaller; 4). The cross-section for the drag force decreases because it is $2b$;  5). The cross-section for the lift force remains constant as $2a$; also, 6). The  $a/b=1$ case essentially is a cylinder with the largest cross-section area; 7). The temperature ratio can affect all these four coefficients for a diffusely reflective ellipse, but not the specularly reflective ellipse. 

There are several observations or conclusions from these four figures. 

a). The drag coefficient for the $a/b=1$ case is the biggest, because it has the largest ellipse perimeter with larger friction force and larger pressure difference between the front and the back side;  as $C_D$ decreases as aspect ratio increases, for both cases; 

b). For each specific $a/b$ ratio, $C_D(0^\circ)$ is the maximum and $C_L(0^\circ)$ is zero, the flow impinges and pushes the ellipse towards the right;

c). $C_D(90^\circ)$ is always zero and $C_L(90^\circ)$ is the maximum and the free stream lifts the ellipse upwards. 

d). Figs. \ref{Fig:slipu_Cp} and \ref{Fig:Cf_Cq} illustrate that the largest value for a local $C_p$ is about 12, while the largest value for the local friction $c_f$ is $1.5$. It means the local friction force is not negligible, and shall be included in the $C_D$ and $C_L$ computations. Many past work completely neglected this friction force factor. This work does include the friction effect - this is another advantage.

e). Fig. \ref{Fig:CD_alpha}  shows that the $C_D(\alpha_0)$ in the diffusely reflective surface case is always larger than the corresponding value for the specular surface scenario. This is because for the diffusely reflective ellipse, the friction forces along the top and bottom sides contribute to $C_D(\alpha_0)$, but there is no friction force in the specular reflective surface. 

f). For $C_L(\alpha_0)$ profiles,  specularly reflective ellipses have more appreciable differences among themselves than a diffusely reflective ellipse; 

g). For a diffusely reflective surface, the aspect ratios do not have much effect on $C_L$; This is different from the drag force along the X-axis, and the cross-section size is $2b$ which continues less as $a/b$ ration increases. For a specular surface, larger $a/b$ ratios contribute less to $C_L$,

h). For $C_M$ and $S_{cc}$, the values for the diffusely reflective ellipse are much larger than those values for their counterparts in the specularly reflective ellipse case;

i). For the cylinder shape with $a/b=1$, $C_M(\alpha_0)=0$ and $S_{cc}(\alpha_0)=0$; 

j). $C_M(\alpha_0=0^\circ) = C_M(\alpha_0=90^\circ) = S_{cc}(\alpha_0=0^\circ) = S_{cc}(\alpha_0=90^\circ)$, due to the symmetries.

k). When $\alpha_0$ is close to $45^\circ$, the moment reaches the maximum; 

l). With large $a/b$ ratios, both $C_M(\alpha_0)$ and $S_{cc}(\alpha_0)$increase monotonically; 

m). The  $S_{cc}(\alpha_0)$ profiles are less symmetric than then $C_M(\alpha_0)$ profiles.  During the past, several formulas (e.g. \cite{Mando, Rosendahl, Yin} )were proposed to describe the $S_cc(\alpha_0)$, for continuum or low Reynolds number flows over an ellipse. Usually these formulas involve simple sine or cosine formats. However, due to the large skewness in the $S_{cc}$ profiles, we do not recommend similar function. 

n). There are great temptations to propose empirical relations for $C_D(\alpha_0)$, $C_L(\alpha_0)$, and $C_M(\alpha_0)$ at this collisionless flow limit, as Sanjeevi \cite{Sanjeevi} suggested for the corresponding coefficients at the continuum limit. For example, it seems plausible to propose simple relations with simple sine or cosine functions, e.g., 
\begin{equation}
     C_D(\alpha_0) = C_D(0^\circ) \cos \alpha_0; C_L(\alpha_0) = C_L(90^\circ) \sin \alpha_0; C_M(\alpha_0) = 2 C_{M} (45^\circ) \sin \alpha_0 \cos \alpha_0.
\end{equation}
However,  the above formulas are not always correct. For example, the $C_D(\alpha_0)$ for a specularly reflective ellipse with the  $a/b=10$ scenario varies almost linearly, not following a cosine function. As such, we do not propose any empirical formulas to the reader.  Instead, with those formulas obtained in this work, those curves can be evaluated accurately and efficiently without numerical simulations. -This is another advantage of this work: we do not need to comprised the accuracy. 

%%%%%%%%%%%%%%%%%%%%%%%%%%%%%%%%%%%%%%%%%%%%%%%%%%%%%%%%%%%%%%%%%
Then we proceed to discuss the speed ratio effect from $S_0$ on those four coefficients. There are several impacts: 8). The expressions for $C_p$ and $C_f$ include $S_0$ factors, exponent and trigonometric functions, also in the denominators which is believed crucial; 9). One significant competing factor from a larger $S_0$ is, there are more high speed molecules impinging at the ram side of the ellipse surface, and less particles can impinge at the backside of the ellipse; 10). As $S_0$ increases, $C_D(\alpha_0)$ decreases - this fact indicates the decay factors such as $1/S_0^2$ and $e^{-S_0^2}$ are important; and 11). As $S_0$ increases, the flows become hypersonic and Newtonian. 

\begin{figure}[ht]
    \begin{minipage}[l] {0.48\textwidth}
      \centering
      \includegraphics[width=3.6in]{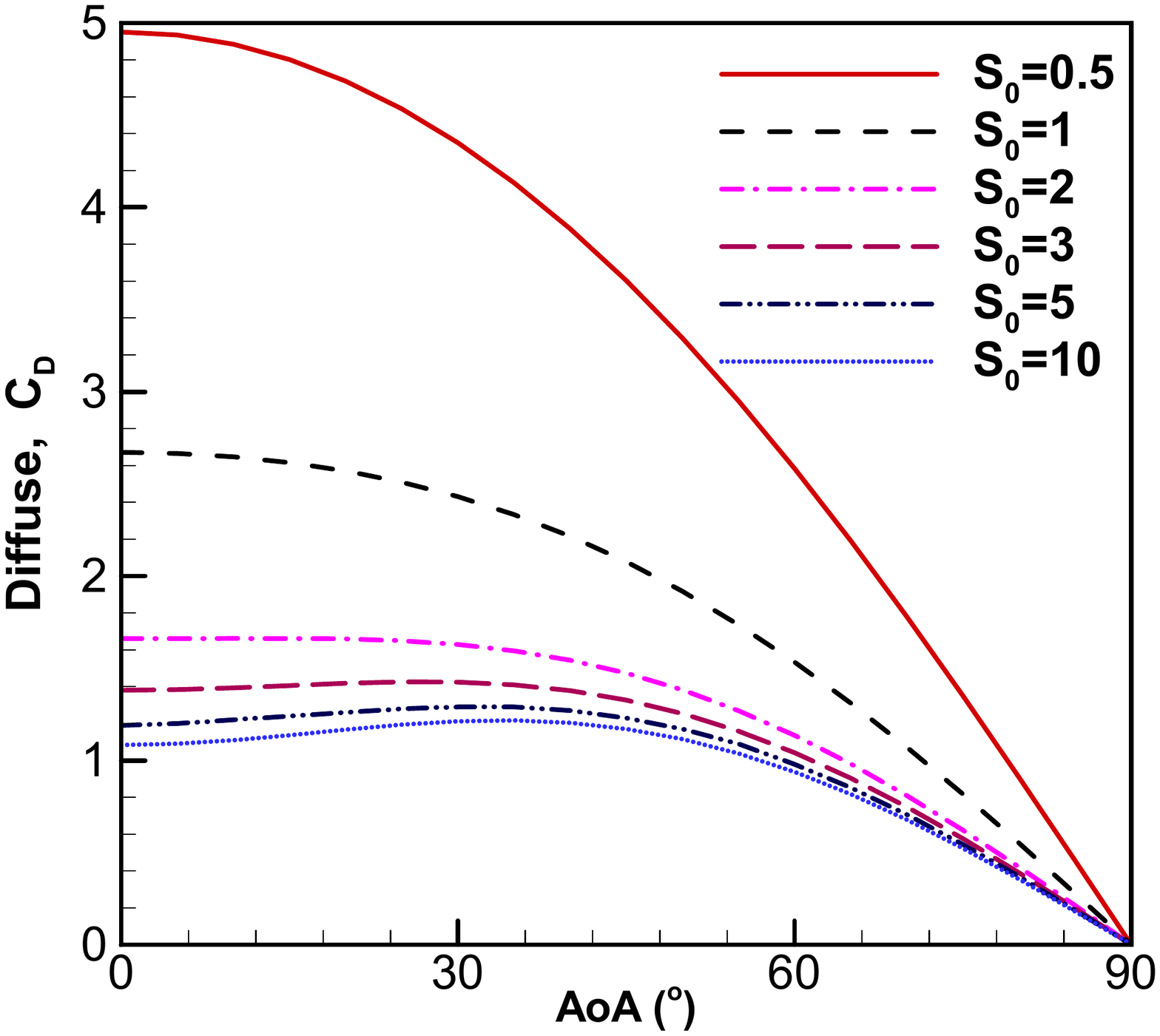}
  \end{minipage}
  \begin{minipage}[l]{0.48\textwidth}
     \centering
      \includegraphics[width=3.6in]{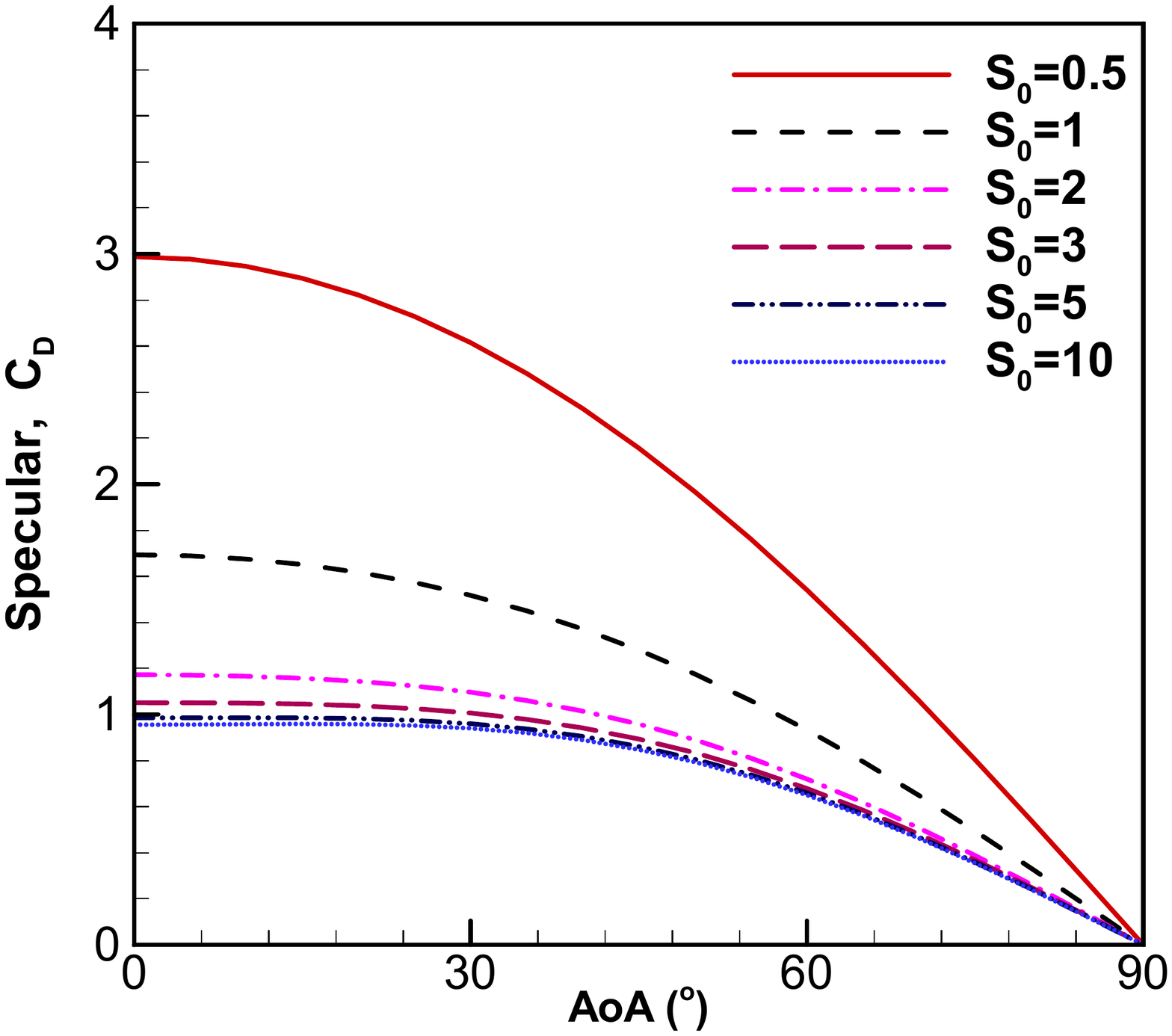}
  \end{minipage}
\caption{$S_0$ effect on $C_D$, diffusely (left) or specularly (right) reflective ellipse. $a/b=2$, and $\epsilon =1.5$. }
\label{Fig:CD_S}
\end{figure}

\begin{figure}[ht]
    \begin{minipage}[l] {0.48\textwidth}
      \centering
      \includegraphics[width=3.6in]{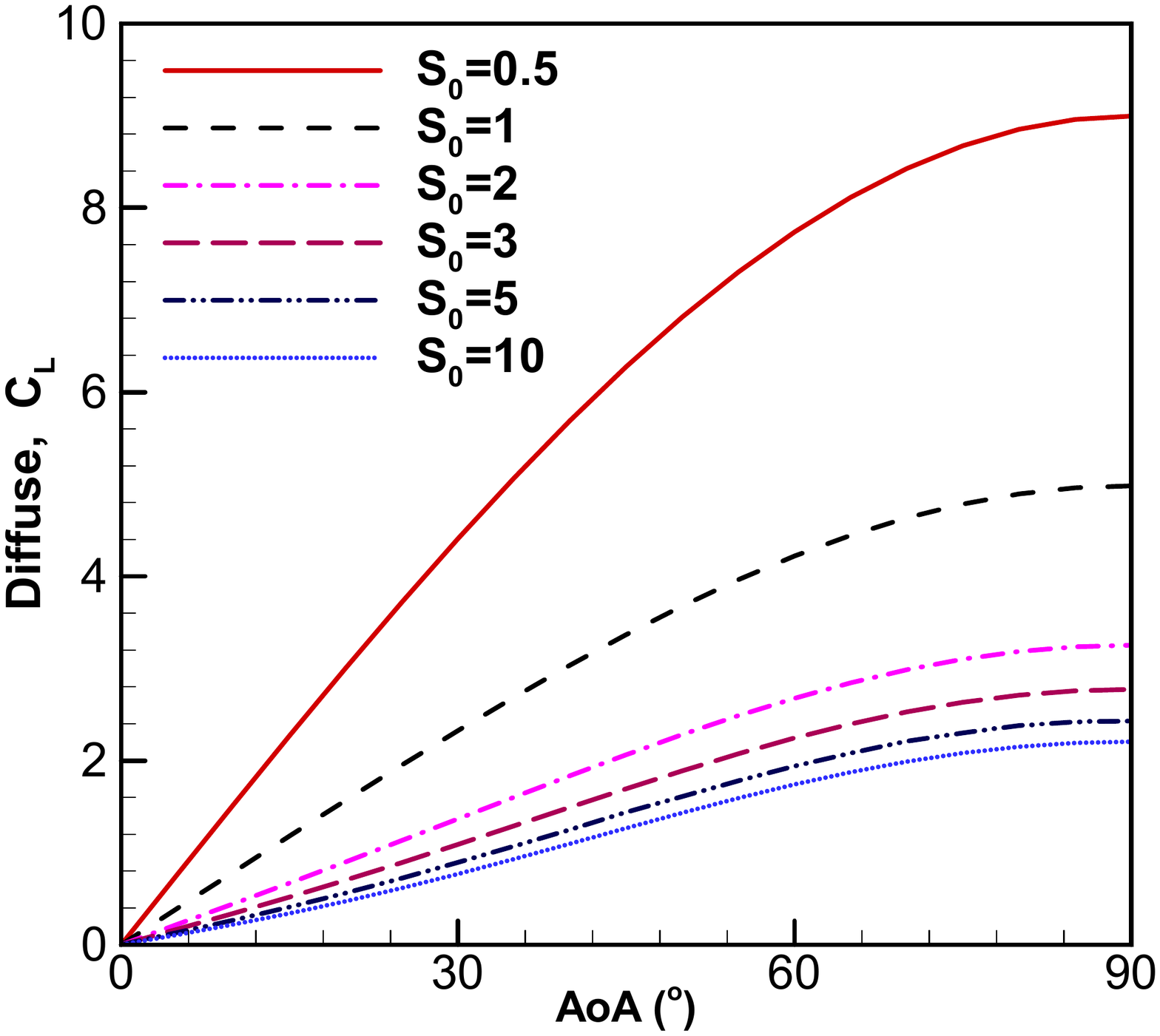}
  \end{minipage}
  \begin{minipage}[l]{0.48\textwidth}
     \centering
      \includegraphics[width=3.6in]{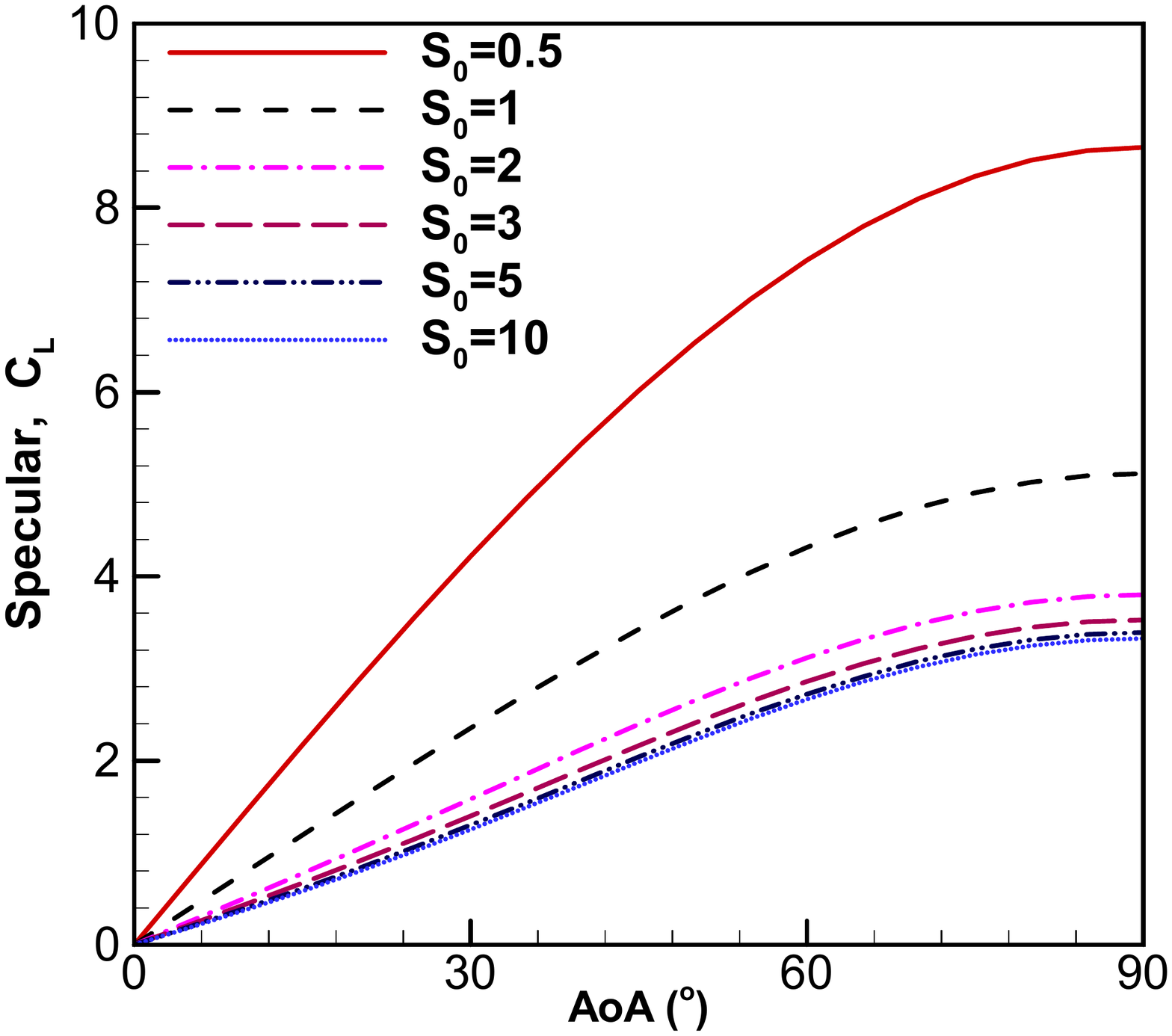}
  \end{minipage}
\caption{$S_0$ effect on specularly (right) reflective ellipse. $a/b=2$, and $\epsilon =1.5$.}
\label{Fig:CL_S}
\end{figure}

\begin{figure}[ht]
    \begin{minipage}[l] {0.48\textwidth}
      \centering
      \includegraphics[width=3.6in]{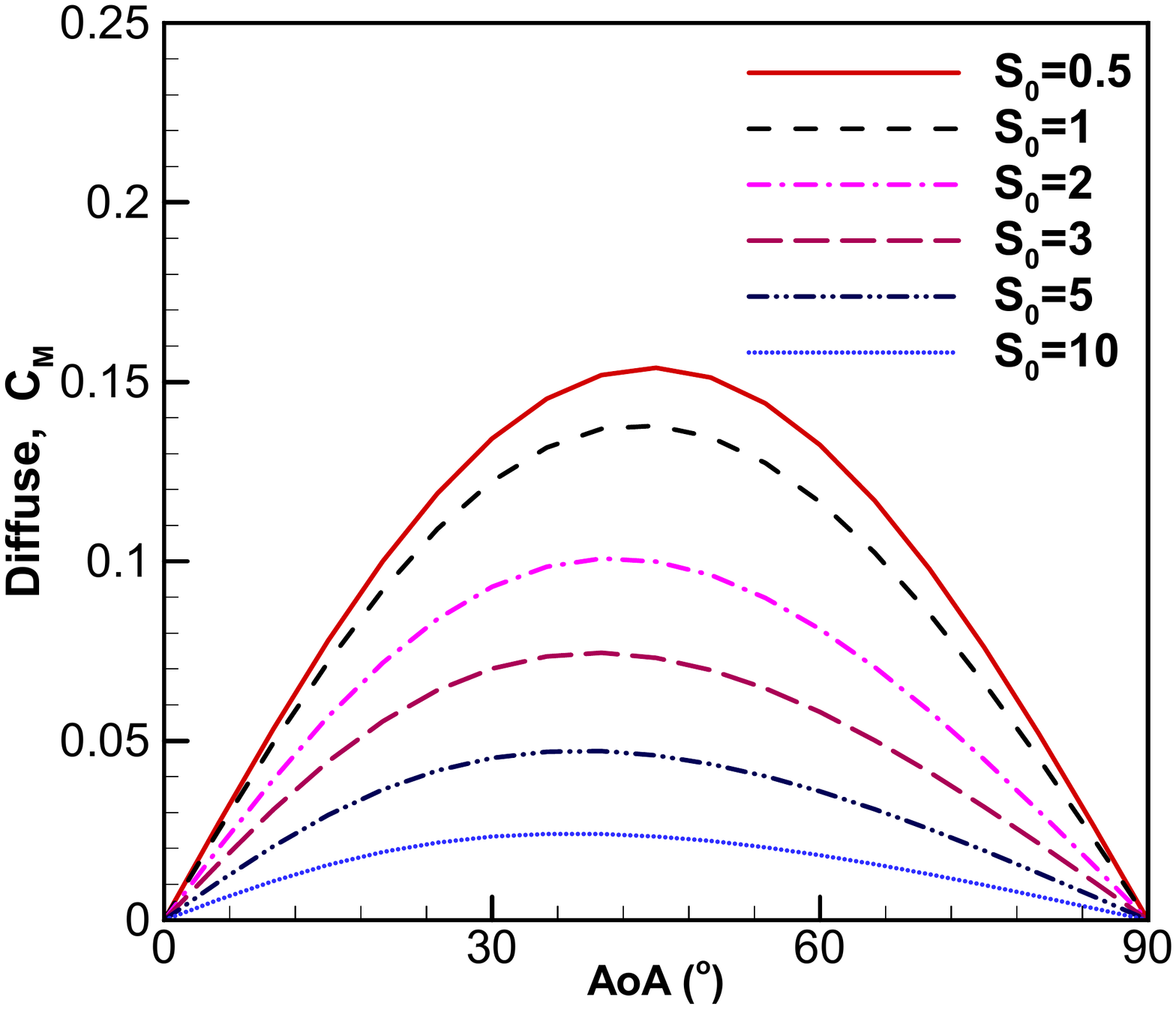}
  \end{minipage}
  \begin{minipage}[l]{0.48\textwidth}
     \centering
      \includegraphics[width=3.6in]{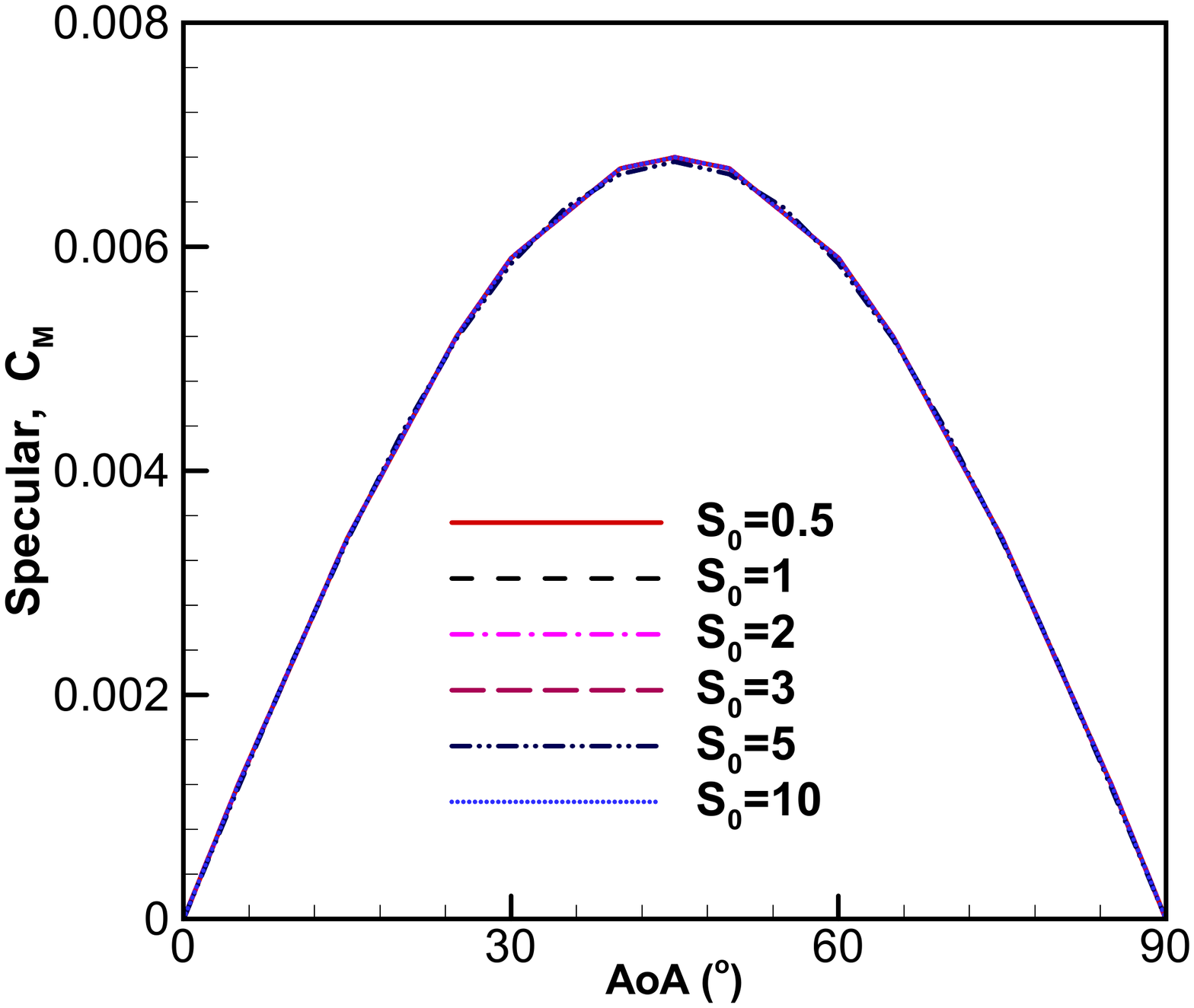}
  \end{minipage}
\caption{$S_0$ effect on $C_M$, diffusely (left) or specularly (right) reflective surface. $a/b=2$, $S_0= 0.1, 0.5, 1.0, 2.0, 3.0$, and $\epsilon =1.5$.}
\label{Fig:CM_S}
\end{figure}

\begin{figure}[ht]
    \begin{minipage}[l] {0.48\textwidth}
      \centering
      \includegraphics[width=3.6in]{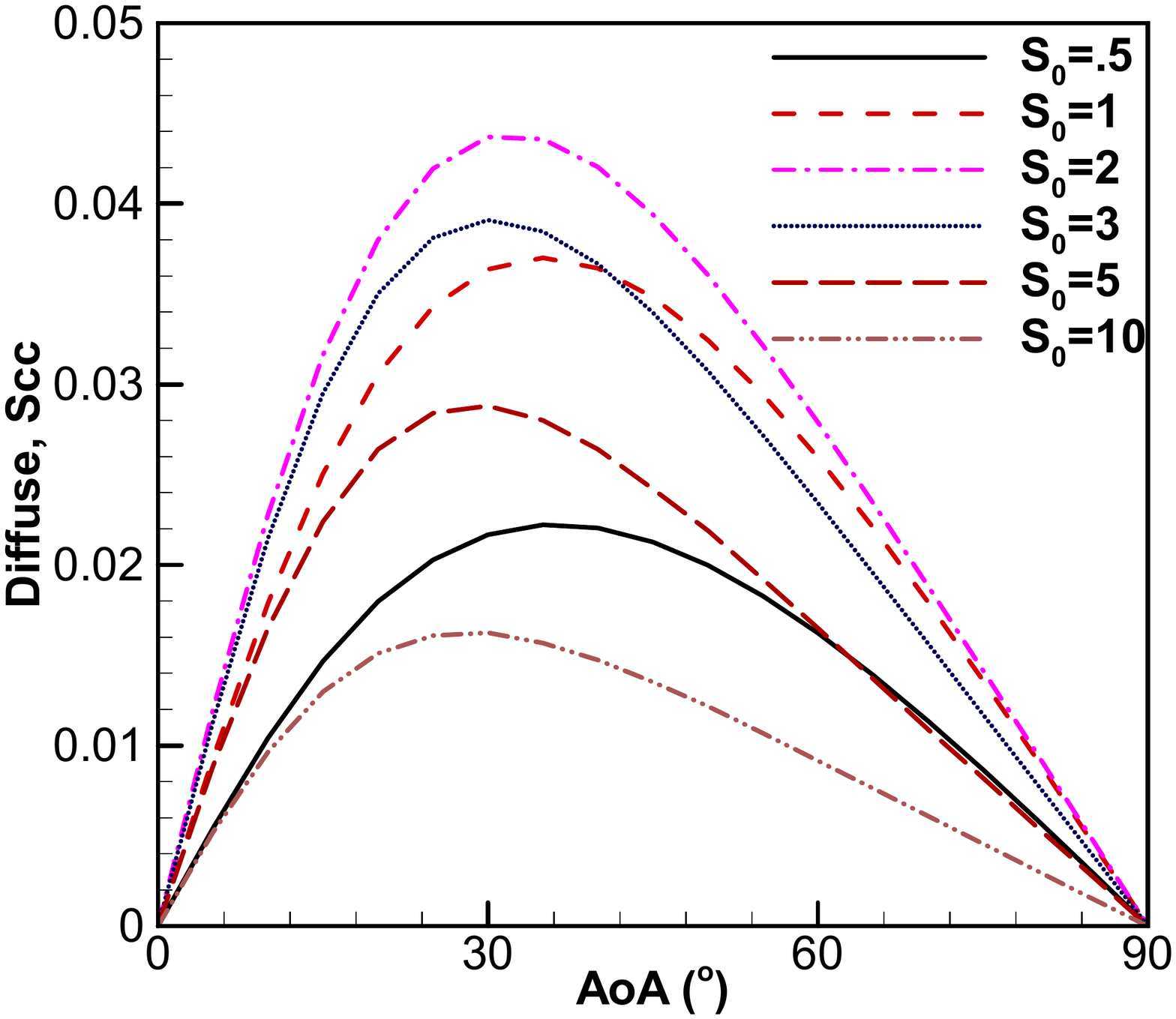}
  \end{minipage}
  \begin{minipage}[l]{0.48\textwidth}
     \centering
     \includegraphics[width=3.6in]{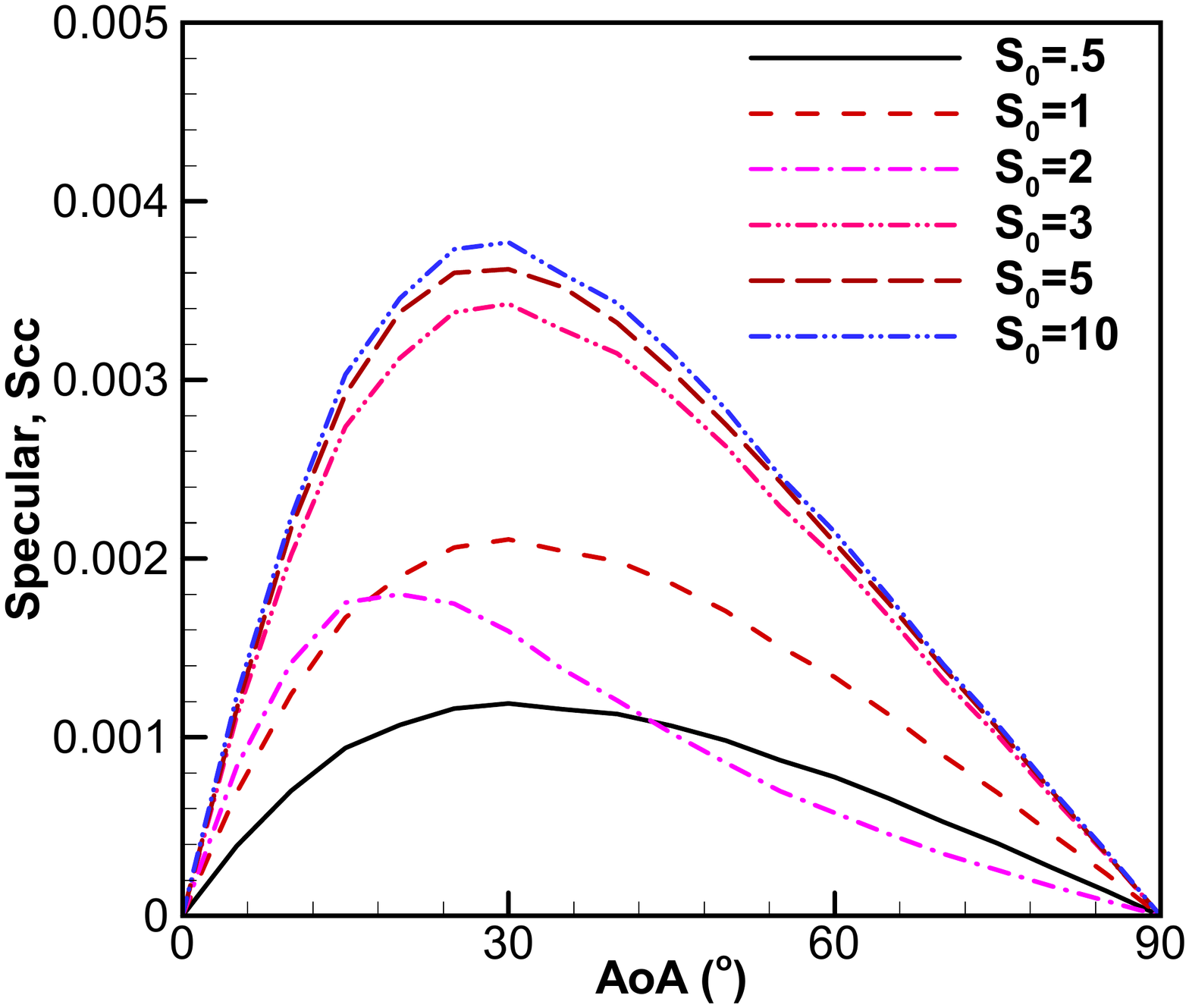}
  \end{minipage}
\caption{$S_0$ effect on $S_{cc}$, diffusely (left) or specularly (right) reflective surface. $a/b=2$, $S_0= 0.1, 0.5, 1.0, 2.0, 3.0$, and $\epsilon =1.5$.}
\label{Fig:speed_scc}
\end{figure}

Figures  \ref{Fig:CD_S}, \ref{Fig:CL_S}, \ref{Fig:CM_S} and \ref{Fig:speed_scc} illustrate the profiles, the ellipse ratio is fixed at $a/b =2$ and $\epsilon =1.5$.  Several observations can be summarized as follows. 

o). The traditional wing theory for continuum flows shows that $C_L(\alpha_0)$ shall not include the speed ratio factor $S_0$ but evidently this is not the situation for collisionless flows;

p). As $S_0=0$, all the curves have the largest profiles, a direct contributor is from the $1/S_)^2$;

q). $C_D(0^\circ)$ reaches the maximum because the friction forces along the top and bottom faces do not cancel. 

r). As $S_0$ increases, the differences among curves become smaller; and the $C_D(\alpha_0)$
and $C_L(\alpha_0)$ profiles become almost identically, correspondingly. This is because hypersonic flows encompasses the ellipse surface more accurately;

s). For $C_M(\alpha_0)$ and $S_{cc}(\alpha_0)$, the values for the diffusely reflective surface are relatively larger than those for the specularly reflective surface;

t). For $S_{cc}(\alpha_0)$, the profiles are do not change monotonically as $S_0$, and a sine function is not a good curve-fitting option. 

u). For the specularly reflective ellipse, $C_M$ becomes almost ir-relevant to the free stream speed ratio $S_0$. This is an interesting result and the reason for this is not clear. 

\begin{figure}[ht]
    \begin{minipage}[l] {0.48\textwidth}
      \centering
      \includegraphics[width=3.6in]{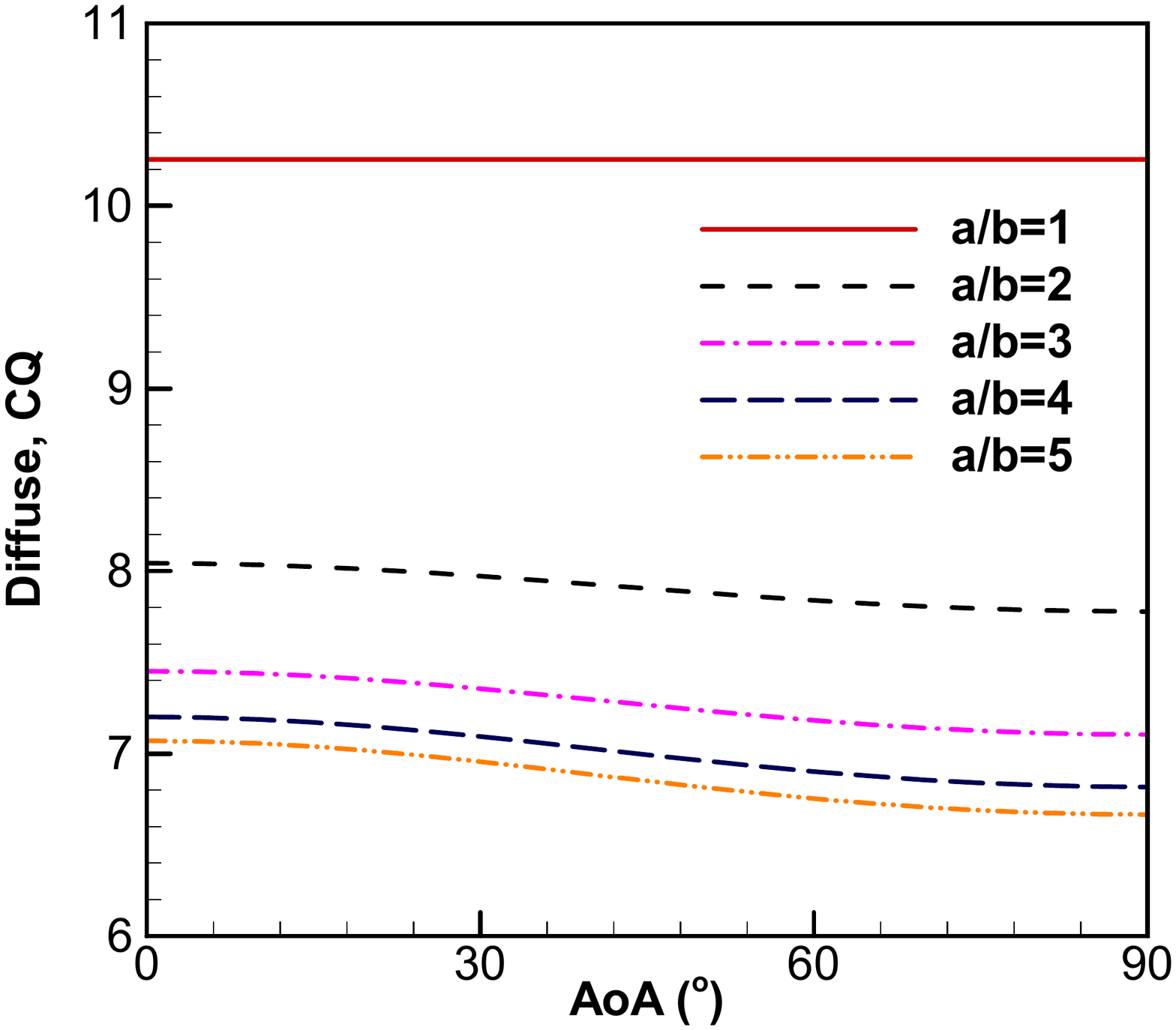}
  \end{minipage}
  \begin{minipage}[l]{0.48\textwidth}
     \centering
      \includegraphics[width=3.6in]{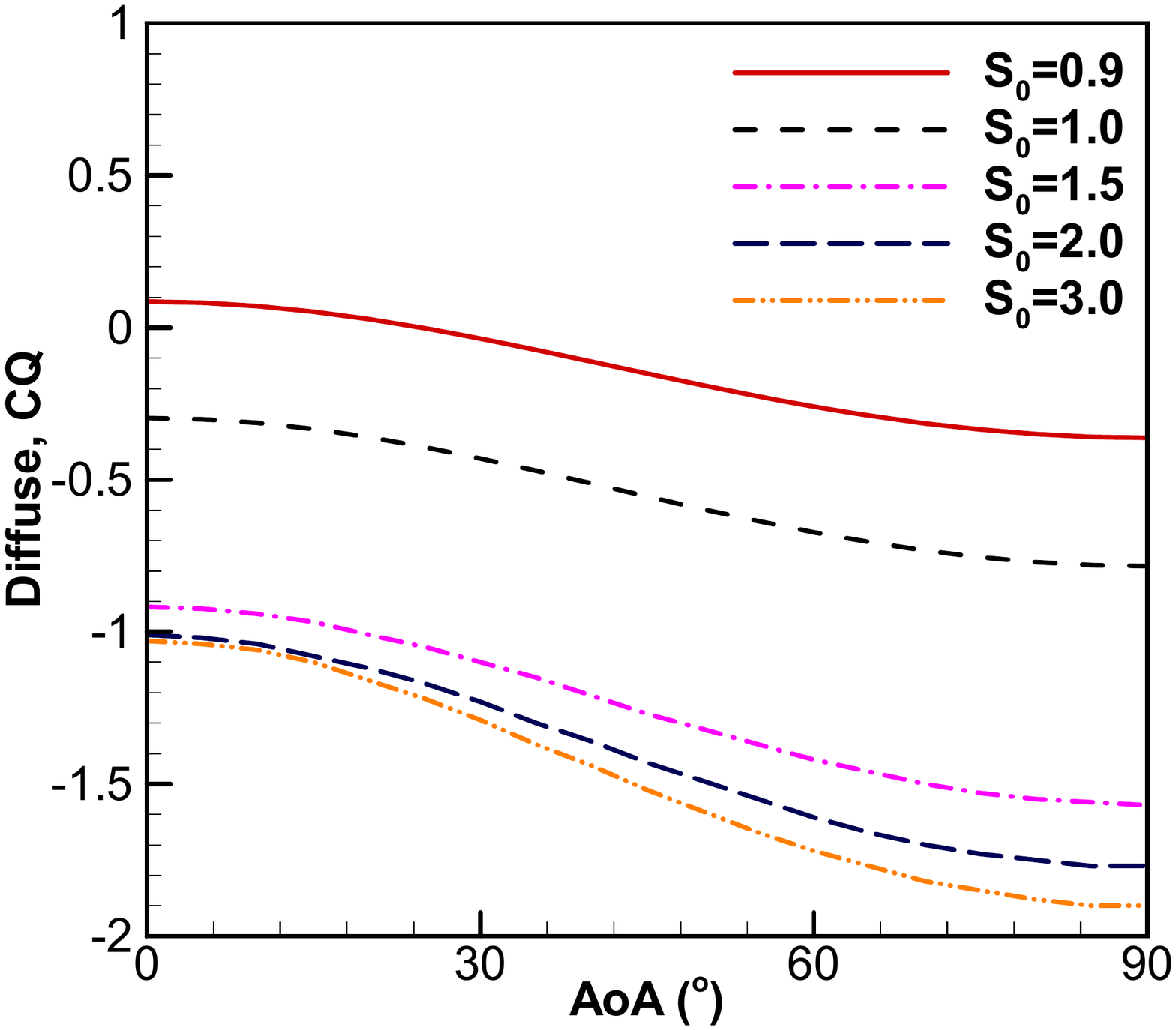}
  \end{minipage}
\caption{Aspect ratio (left) and $S_0$ (right) effect on $C_Q$, diffusely reflective ellipse, $\epsilon =1.5.$}
\label{Fig:CQ_ratio_S}
\end{figure}

The total heat load coefficient along a diffusely reflective surface can also be integrated out as:
\begin{equation}
       C_Q(\alpha_0) = \frac{1}{2a} \int_0^{2\pi}   C_q(\theta) L(\theta) d\theta.
\label{eqn:CQ}
\end{equation}
Figure \ref{Fig:CQ_ratio_S} shows the aspect ratio (left) or the speed ratio (right) effects on $C_Q(\alpha_0)$.  There are a few observations. 

w). For the cylinder case, with $a/b=1$, $C_Q(\alpha_0)$ does not change with $\alpha_0$;

x). This figure shows with a fixed $S_0$, as the $a/b$ ratio increases, the integrated $C_Q$ becomes smaller;

y). The right side in Fig. \ref{Fig:CQ_ratio_S} shows the $S_0$ effects on $C_Q(\alpha_0)$ with $a/b=2$. Note for the case $S_0 =0.9$, within a small $\alpha_0$, $C_Q(\alpha_0)$ is positive, but as $\alpha_0$ increases, $C_Q$ becomes negative. This means the net heat flux changes its transfer directions;  and 

z). With a sufficiently large $S_0$, the effect from $S_0$ change is insignificant, because the dominating $1/S_0^3$ and $e^{-S_0^2}$ factors do not change much. With an increasing $S_0$, the $C_Q$ profile approaches an asymptote distribution, and the collisionless gas flows become hypersonic and Newtonian.
%%%%%%%%%%%%%%%%%%%%%%%%%%%%%%%%%%%%%%%%%%%%%%%%%%%%%%%%%%%%%%%%%

%%%%%%%%%%%%%%%%%%%%%%%%%%%%%%%%%%%%%%%%%%%%%%%%%%%%%
\section{FLOWFIELD AROUND AN ELLIPSE}
\label{sec:field}
In addition to the surface properties, external flowfields around an ellipse can offer some insights. They can also provide further validations because they are more complex and it is challenging to have good agreement with simulation results.  To provide a set of complete work, the flowfield property expressions are also developed, validated and presented in this work. 

The investigation method is still based on the VDFs, velocity phases, and the gaskinetic theory. Two past papers\cite{cai_jsr, impingement} adopted the same approach and may aid understanding. 
Figure \ref{Fig:flowfield_illustration} sketches the flowfields around a diffusely (left) and a specularly reflective (right) ellipse. The corresponding velocity phases for Point $P(X,Y)$ are also included in this figure.  

\subsection{Diffusely Reflective Ellipse}
\begin{figure}[ht]
    \begin{minipage}[l] {0.48\textwidth}
      \centering
      \includegraphics[trim=0 70 0 20, clip, height=2.4 in, width=3in]{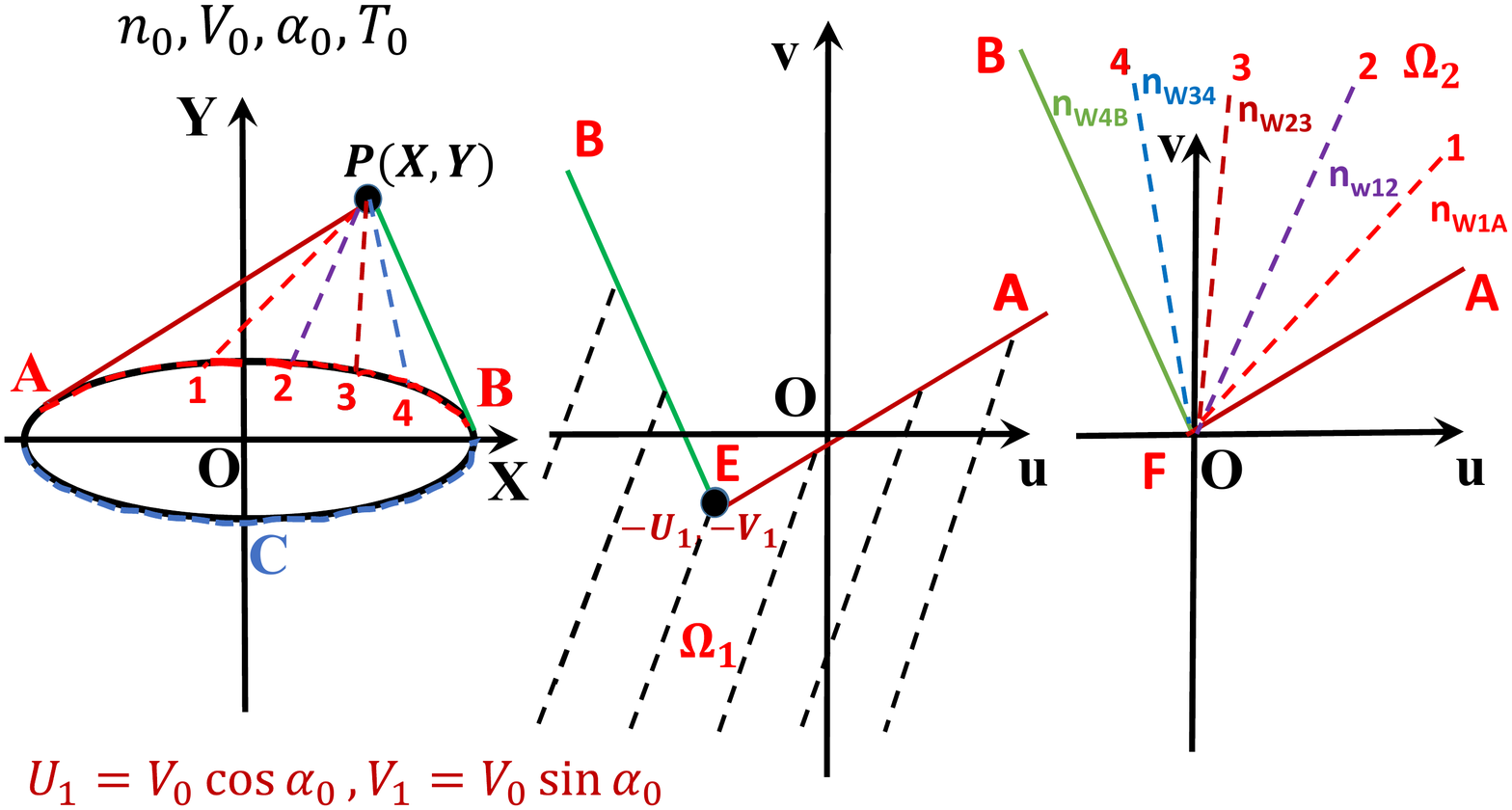}
  \end{minipage}
  \begin{minipage}[l]{0.48\textwidth}
     \centering
      \includegraphics[trim=0 70 0 20, clip, height=2.4in, width=3in]{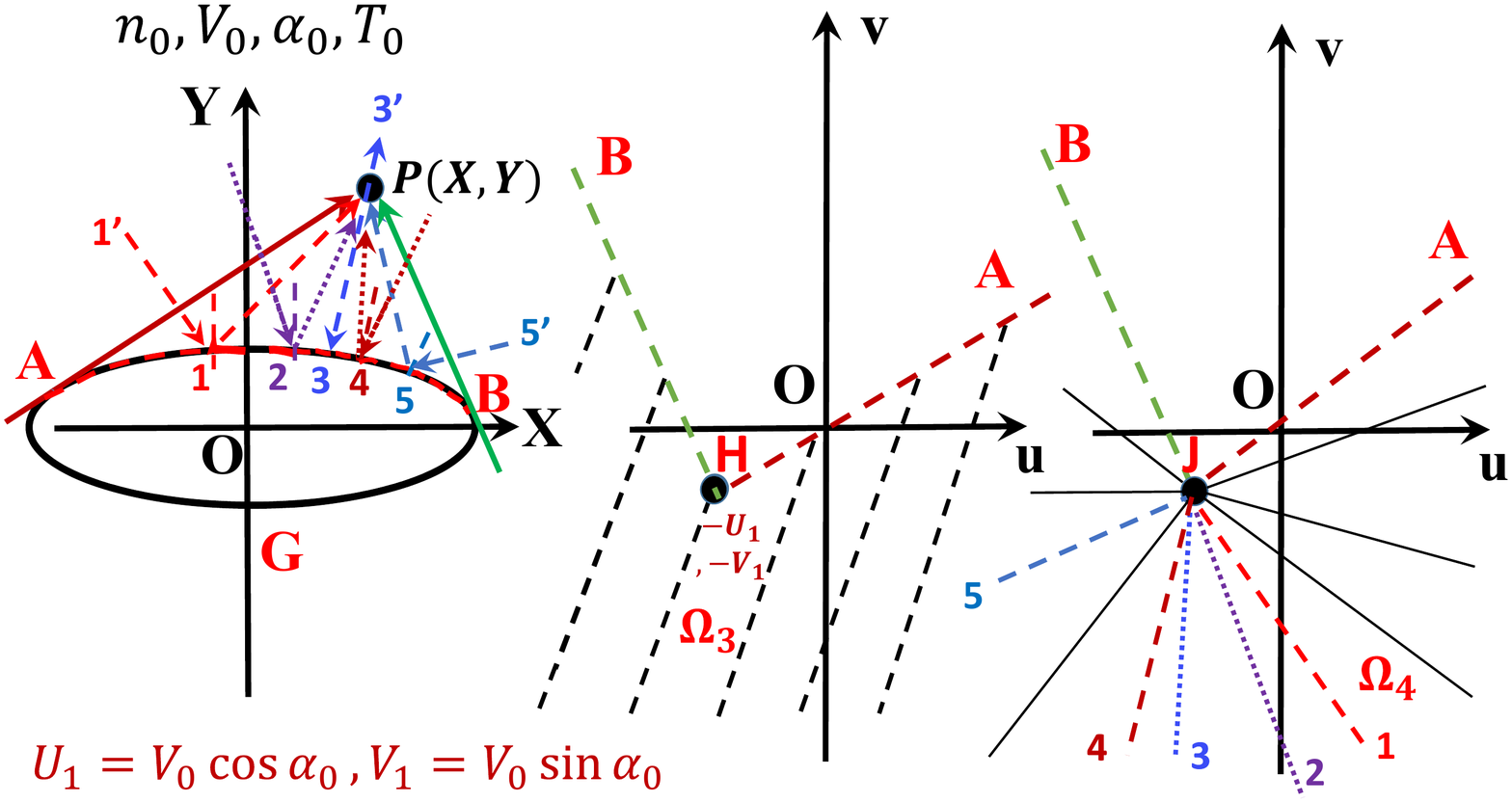}
  \end{minipage}
\caption{Velocity phases for a flowfield point $P(X,Y)$. Left: diffusely reflective surface; Right: specularly reflective surface.}
\label{Fig:flowfield_illustration}
\end{figure}

The left side of Fig. \ref{Fig:flowfield_illustration} shows the diffusely reflective ellipse scenario. There are two sources that can contribute to the properties at Point $P(X,Y)$: the free stream, and the ellipse surface. The corresponding velocity phases are shown next to the flowfield picture. In the flowfield and velocity phases, there are lines with relations and they are drawn with the same color. For example, $AP \parallel AE \parallel AF$, and they are drawn with the brick color; $BP \parallel BE \parallel BF$ and drawn in green. The shadowed region $\Omega_1$ is formed by lines $AE$ and $BE$, and it is the velocity phase for the free stream. The related VDF for this region is Maxwellian and it is characterized by $n_0$, $T_0$ and $V_0$.  $\Omega_2$ is the velocity phase for those molecules bouncing off Point $P(X,Y)$. The VDF for $\Omega_2$ is another Maxwellian VDF characterized by $T_w$ and a virtual number density $n_w(\theta)$. However, different from $\Omega_1$ which has a constant $n_0$, the $n_w(\theta)$ in $\Omega_2$ is not constant and its value shall be determined by using the non-penetration boundary condition, i.e. Eqn. \ref{eqn:nw}, in a segment by segment fashion; and on each segment, $n_w$ can be considered constant. For example, the virtual number density  along arc $12$ is marked as $n_{w12}$ in the velocity phase $\Omega_2$. 

To compute flowfield properties around a diffusely reflective ellipse, one convenient approach is to divided the ellipse surface into 360 small segments or panels, and add their contribution to the flowfield properties segment by segment. For example, the arc visible from Point $P(X,Y)$ can be divided into segments $A1$, $12$, $23$, $34$, and $4B$ in $\Omega_2$, and their contribution shall be considered independently. This approach has similar spirits as the popular panel method in the traditional Computational Fluid Dynamics (CFD) for incompressible potential flows. The virtual density  along the major arc $ACB$ does not contribute to the density at Point $P(X,Y)$ because those small segments are invisible from Point $P(X,Y)$. As such, the major arc $ACB$ is skipped during the computations. 

By using the gaskinetic theory, the two Maxwellian VDFs, and velocity phases $\Omega_1$ and $\Omega_2$, the following expressions for the flowfield density are obtained: 
\begin{equation}
       \frac{n_d(X,Y)}{n_0} =\frac{1}{2 \pi} \int_{\theta_1}^{\theta_2} \frac{n_w(\theta)}{n_0} d\theta +
          1 - \frac{e^{-S_0^2}}{\pi}  \int_{\theta_1}^{\theta_2} e^{K^2 } B^*(K) d\theta,
\label{Eqn:diffuse_n}
\end{equation}
where $\theta_1$ and $\theta_2$ are the two angles subtended by the minor arc $ACB$ visible at the flowfield Point $P(X,Y)$. For example, in Fig. \ref{Fig:flowfield_illustration}, from Point $P(X,Y)$, arc $\widehat{A1234B}$ is visible, $\theta_1 =\angle Aou$ and $\theta_2 = \angle Bou$. The first term in Eqn. \ref{Eqn:diffuse_n} represents the contribution from the minor arc at the ellipse surface, and the remaining two terms represent the free stream contribution.

Following the same steps, the formulas for the velocity components and temperature are obtained: 
\begin{equation}
       \frac{U_d(X,Y)}{\sqrt{2RT_0}} = \frac{\sqrt{\epsilon} }{4 \sqrt{\pi} } \frac{1}{n_d} \int_{\theta_1}^{\theta_2}      n_w(\theta) \cos \theta d\theta  + S_0 \cos \alpha_0 - \frac{e^{-S_0^2}}{\pi} \frac{n_0}{n_d}  \int_{\theta_1}^{\theta_2} e^{K^2 } \cos \theta C^*(K) d\theta,
\label{Eqn:diffuse_U}
\end{equation}
\begin{equation}
       \frac{V_d(X,Y)}{\sqrt{2RT_0}} =  \frac{\sqrt{\epsilon}  }{4 \sqrt{\pi} n_d }    \int_{\theta_1}^{\theta_2}      n_w(\theta) \sin \theta d\theta   + S_0 \sin \alpha_0 - \frac{e^{-S_0^2}}{\pi} \frac{n_0}{n_d}  \int_{\theta_1}^{\theta_2} e^{K^2 } \sin \theta C^*(K) d\theta,
\label{Eqn:diffuse_V}
\end{equation}
\begin{equation}
       \frac{T_d(X,Y)}{T_0} =  \frac{\epsilon}{3\pi} \frac{1}{n_d} \int_{\theta_1}^{\theta_2}  n_w(\theta) d\theta   + 1 - \frac{e^{-S_0^2}}{3\pi} \frac{n_0}{n_d }  \int_{\theta_1}^{\theta_2} e^{K^2}\left( D^*(k)+ \frac{B^*(K)}{2} \right)d\theta - \frac{U_d^2 + V_d^2 }{3RT_0} \frac{n_0}{n_d}.
\label{Eqn:diffuse_T}
\end{equation}

\subsection{Specularly reflective ellipse}
Figure \ref{Fig:flowfield_illustration} sketches the flowfield around a specularly reflective ellipse and the velocity phases for Point $P(X,Y)$. The contribution from the free stream is represented in Region $\Omega_3$, which is identical to $\Omega_1$ for the diffusely reflective ellipse scenario. The related computation scheme and results for the free stream contribution is identical to those for the diffusely reflective ellipse scenario.  As to the contribution from the specularly reflective surface visible from Point $P(X,Y)$, i.e., the minor arc $ACB$, the treatments are genuinely challenging. In one past paper\cite{cai_jsr} for free molecular flows over a specularly reflective sphere or a cylinder, a coordinate rotation is first performed and it can significantly simplify the derivations. After proper computations, an inverse rotation is performed. However, such treatment is more challenging for this complex ellipse situation.  One relatively straightforward approach is to loop over those segments visible from Point $P(X,Y)$, i.e., the minor arc $AB$. The contributing molecules can be traced back to area $\Omega_4$, which is represented with $\widehat{B54321A}$ and this area is identical to $\Omega_3$. There are also parallel relations, for example, $AP \parallel AH \parallel AJ$, $BP \parallel BH \parallel BJ$, $1'1 \parallel 1J$,  $33' \parallel 3J$, and  $ 55' \parallel 5J$.  When we loop over each surface segment or panel visible from Point $P(X,Y)$, the normal direction at that segment is first computed, and a mirror mapping action between $\Omega_2$ and $\Omega_4$ is then performed to find the corresponding specific ray in Region $\Omega_4$.  Along different rays in $\Omega_4$, e.g., $1J$ and $3J$, the corresponding incoming fluxes are different.  In this work, this approach is adopted because it is more straightforward than the rotation approach \cite{cai_jsr}.  
The final expressions for the flowfield density, velocity components, and temperature are similar to Eqns. \ref{Eqn:diffuse_n}, \ref{Eqn:diffuse_U}, \ref{Eqn:diffuse_V} and \ref{Eqn:diffuse_T}. They are:
\begin{equation}
       \frac{n_s(X,Y)}{n_0} = \Pi_n +
          1 - \frac{e^{-S_0^2}}{\pi}  \int_{\theta_1}^{\theta_2} e^{K^2 } B^*(K) d\theta,
\label{Eqn:specular_n}
\end{equation}
\begin{equation}
       \frac{U_s(X,Y)}{\sqrt{2RT_0}} = \Pi_U + S_0 \cos \alpha_0 - \frac{e^{-S_0^2}}{\pi} \frac{n_0}{n_s}  \int_{\theta_1}^{\theta_2} e^{K^2 } \cos \theta C^*(K) d\theta,
\label{Eqn:specular_U}
\end{equation}
\begin{equation}
       \frac{V_s(X,Y)}{\sqrt{2RT_0}} =  \Pi_V   + S_0 \sin \alpha_0 - \frac{e^{-S_0^2}}{\pi} \frac{n_0}{n_s}  \int_{\theta_1}^{\theta_2 } e^{K^2 } \sin \theta C^*(K) d\theta,
\label{Eqn:specular_V}
\end{equation}
\begin{equation}
       \frac{T_s(X,Y)}{T_0} =  \Pi_T  + 1 - \frac{e^{-S_0^2}}{3\pi} \frac{n_0}{ n_s }  \int_{\theta_1}^{\theta_2} e^{K^2} \left( D^*(K)+ \frac{B^*(K)}{2} \right)d\theta - \frac{U_s^2 + V_s^2 }{3RT_0} \frac{n_0}{n_s},
\label{Eqn:specular_T}
\end{equation}
where $\Pi_n$, $\Pi_U$, $\Pi_V$ and $\Pi_T$ represent the contribution from those segments or panels visible from Point $P(X,Y)$. Due to the specular reflection relation at each segment point, it is difficult to develop their explicit formulas by using $\Omega_2$ in Fig. \ref{Fig:flowfield_illustration}. Instead, for each segment or panel on the surface, first the specific angle within $\Omega_2$ is identified, then the mapping relation involving the normal direction is adopted to find the mirrored angle within $\Omega_4$. The specific angle within $\Omega_4$ is used to compute the incoming number fluxes,  and the momentum contribution to $U(X,Y)$ and $V(X,Y)$ shall be converted with the mirror relation. 

\begin{figure}[ht]
    \begin{minipage}[l] {0.48\textwidth}
      \centering
 \includegraphics[width=6.0in]{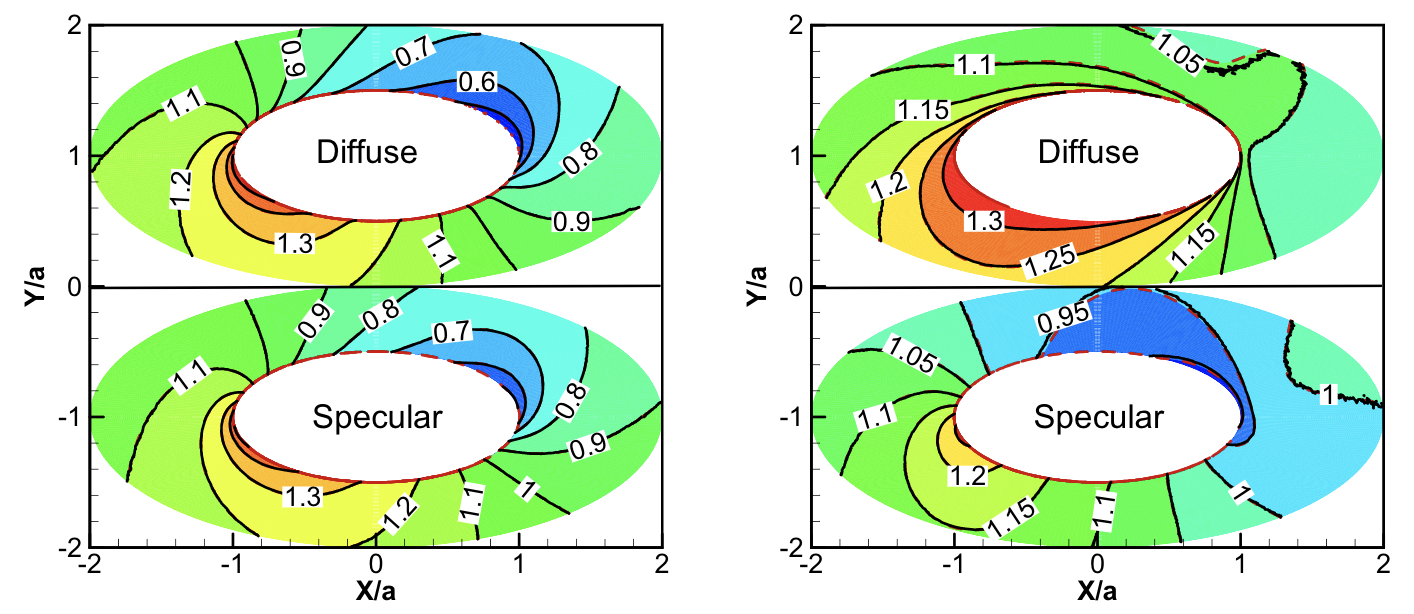}
  \end{minipage}
\caption{Normalized density $n(X,Y)/n_0$ (left) and temperature $T(X,Y)/T_0$ (right) around a diffusely (top) or a specularly (bottom) reflective ellipse, $a/b=2$, $\alpha_0 = 30^\circ$, $S_0=0.5$, and $\epsilon =1.5$. Dashed: analytical; Solid: DSMC. }
\label{Fig:field_nT}
\end{figure}

%%%%%%%%  bellow is the  flowfield results %%%%
Figure \ref{Fig:field_nT} compares the number density (left) and translational temperature (right) contours around a diffusely (top) and a specularly reflective (bottom) ellipse. It is evident that the corresponding flowfield patterns are similar. The difference in density is minor, and the temperature fields have more appreciable differences. The temperatures around the diffusely reflective surface is relatively higher than the specular surface because $\epsilon = T_w/T_0 =1.5$, the free stream cold gas shall eventually warm up to the surface temperature.  For a specularly reflective surface, the surface temperature does no affect those reflected particles. 

\begin{figure}[ht]
    \begin{minipage}[l] {0.48\textwidth}
      \centering
     \includegraphics[width=6.0in]{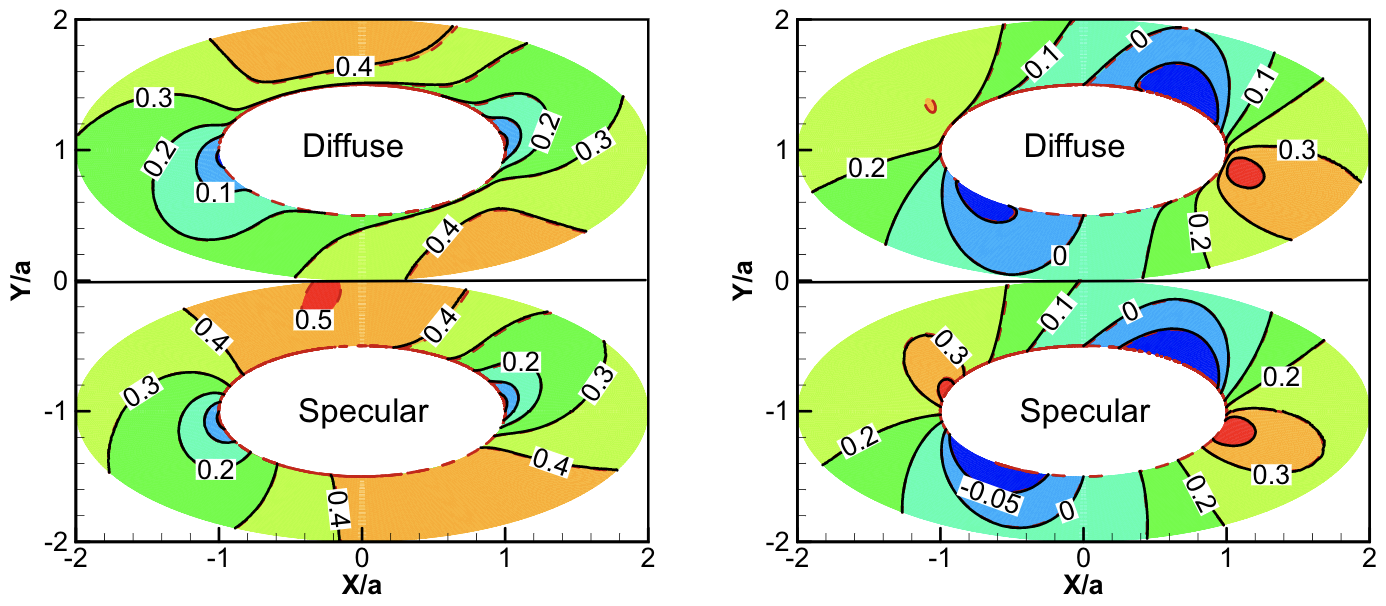}
  \end{minipage}
\caption{Normalized flowfield velocity components, $U(X,Y)/\sqrt{2RT_0}$ (left) and  $V(X,Y)/\sqrt{2RT_0}$ (right) around a diffusely (top) or a specularly (bottom) reflective ellipse, $a/b=2$, $\alpha_0 = 30^\circ$, $S_0=0.5$, and $\epsilon =1.5$. Dashed: analytical; Solid: DSMC. }
\label{Fig:field_UV}
\end{figure}

Figure \ref{Fig:field_UV} shows the velocity components around a diffusely or a specularly reflective ellipse. As shown the flowfield patterns are complex, and there are much appreciable differences between the diffusely and specularly reflective ellipses. However, the good agreements between the simulation and exact results indicate the exact flowfield solutions are correct. 

%\section{DISCUSSIONS} 
%\label{sec:discussion}
%{\color{red}Comparing with the past work, and show differences.. to be added later...}  

\section{CONCLUSIONS}
\label{sec:conclusion}
This paper reported recent work on free molecular flows over a diffusely or specularly reflective ellipse. Gaskinetic methods are adopted with proper constructions of velocity phases and VDFs. Expressions for local surface properties, such as velocity slips, pressure, friction and heat flux are obtained and validated with DSMC simulations. Coefficients for the total life and drag forces, moment, and mass-center to force-center distance are obtained, and effect from various factors were discussed numerically in systematic and effective manners. Different from many past work, the local friction force is included in the total drag and lift force computations. Flowfield properties are also obtained by carefully constructing the velocity phases and VDFs, and applying the gaskinetic theory. In the flowfield computations, the ellipse surfaces are treated as small segments or panels. DSMC simulations are performed and they validate these analytical surface and flowfield solutions. The essentially identical agreement between the simulation results and exact expressions indicates that this work is solid. 

These results may offer baseline solutions with several physical factors, i.e, the aspect ratio, speed ratio, and the temperature ratio (for a diffusely reflective ellipse only). Those surface and flowfield properties need numerical evaluations which usually can be finished in one minute or even seconds by using a personal computer. By comparison, DSMC simulations may need days. As such, the evaluation cost can be significantly reduced. In addition, the variations due to parameter changes can be promptly investigated in a systematic manner and the results can effectively guide parameter optimizations. The expressions for surface properties are useful, and because the flow is collisionless, these expressions may serve as baseline references for quick engineering estimations for less rarefied gaseous flow over an ellipse.    

\section*{APPENDIX}
Several useful integrals and definitions for $A^*$, $B^*$, $C^*$, $D^*$, and $E^*$:
\begin{equation}
      \int_{\pm a}^{\infty } e^{- r^2}  dr = \frac{\sqrt{\pi}}{2}  \left[ 1 \mp \mbox{erf}(  a)  \right], 
%\end{equation}
%\begin{equation}
      \int_{\pm a}^{\infty }  r e^{- r^2}  dr =  \frac{1}{2} e^{- a^2},  
\end{equation}
\begin{equation}
      \int_{\pm a}^{\infty }  r^2 e^{- r^2} dr  =     \frac{\sqrt{\pi} }  {4} \left[ 1 \mp \mbox{erf}(  a)  \right] \pm \frac{a}{2} e^{- a^2},
%\end{equation}
%\begin{equation}
      \int_{\pm a}^{\infty }  r^3 e^{- r^2} dr  =  \frac{1+  a^2}{2} e^{- a^2},
\end{equation}
\begin{equation}
      \int_{\pm a}^{\infty }  r^4 e^{- r^2} dr  =  \frac{3\sqrt{\pi}}{8}   \left[ 1 \mp \mbox{erf}( a)  \right]  \pm   \left( \frac{3a}{4}+ \frac{a^3}{2} \right) e^{- a^2},
%\end{equation}
%\begin{equation}
      \int_{\pm a}^{\infty }  r^5 e^{- r^2} dr  =  \left( 1+ a^2 + \frac{a^4}{2}   \right) e^{- a^2}.
\end{equation}

%old integrals
%\begin{equation}
%      \int_{0}^{\infty}  e^{-\beta (r-A)^2}dr   = \frac{1}{2} \sqrt{\frac{\pi}{\beta}} \left[ 1 + \mbox{erf}( \sqrt{\beta} A)  \right] \equiv A^*(A)
%\end{equation}
%\begin{equation}
%      \int_{0}^{\infty }  r e^{-\beta (r-A)^2}dr   = \left[ 1 + \mbox{erf}( \sqrt{\beta} A)  \right] \frac{A}{2}\sqrt{ \frac{\pi}{\beta}} +\frac{1}{2\beta} e^{-\beta A^2} \equiv B^*(A)
%\end{equation}
%\begin{equation}
%      \int_{0}^{\infty }  r^2 e^{-\beta (r-A)^2}dr   = \left[ 1 + \mbox{erf}( \sqrt{\beta} A)  \right] \left[\frac{A^2}{2} + \frac{1}{4\beta} \right]\sqrt{ \frac{\pi}{\beta}} +\frac{A}{2\beta} e^{-\beta A^2} \equiv C^*(A) 
%\end{equation}

%\begin{equation}
%      \int_0^{\infty}  r^3 e^{-\beta (r-A)^2} dr   = \left[ 1 + \mbox{erf}( \sqrt{\beta} A)  \right] \left[\frac{A^3}{2} + \frac{3A}{4\beta} \right]\sqrt{ \frac{\pi}{\beta}} + \left(\frac{1}{2\beta^2} + \frac{A^2}{2\beta} \right)e^{-\beta A^2} \equiv D^*(A)
%\end{equation}

%\begin{equation}
%      \int_0^{\infty}  r^4 e^{-\beta (r-A)^2} dr   = \left[ 1 + \mbox{erf}( \sqrt{\beta} A)  \right] \left[  \frac{A^4}{2} + \frac{3A^2}{2\beta} + \frac{ 3}{ 8 \beta^2} \right]\sqrt{ \frac{\pi}{\beta}} + \left(\frac{A^3}{2\beta} + \frac{5A}{8\beta^2} \right)e^{-\beta A^2} \equiv E^*(A)
%\end{equation}
%new integrals

\begin{equation}
      \int_{0}^{\infty}  e^{- (r-a)^2}dr   =   \left[ 1 + \mbox{erf}(  a)  \right]   \frac{\sqrt{\pi}}{2} \equiv A^*(a),
\end{equation}
\begin{equation}
      \int_{0}^{\infty}  re^{-(r-a)^2}dr   = \left[ 1 + \mbox{erf}( a)  \right] \frac{\sqrt{\pi}}{2} a +\frac{1}{2} e^{-a^2} \equiv B^*(a),
\end{equation}
\begin{equation}
      \int_{0}^{\infty}  r^2 e^{-(r-a)^2}dr   = \left[ 1 + \mbox{erf}(a)  \right] \left[\frac{a^2}{2} + \frac{1}{4} \right]\sqrt{ \pi} +\frac{a}{2} e^{-a^2} \equiv C^*(a),
\end{equation}
\begin{equation}
      \int_0^{\infty}  r^3 e^{-(r-a)^2} dr   = \left[ 1 + \mbox{erf}( a)  \right] \left[\frac{a^3}{2} + \frac{3a}{4} \right]\sqrt{ \pi} + \left(\frac{1}{2} + \frac{a^2}{2} \right)e^{-a^2} \equiv D^*(a),
\end{equation}
\begin{equation}
      \int_0^{\infty}  r^4 e^{-(r-a)^2} dr   = \left[ 1 + \mbox{erf}(  a)  \right] \left[  \frac{a^4}{2} + \frac{3a^2}{2} + \frac{ 3}{ 8} \right]\sqrt{ \pi} + \left(\frac{a^3}{2} + \frac{5a}{4} \right)e^{-a^2} \equiv E^*(a).
\end{equation}

\reftitle{REFERENCES}

%%%%%%%%%%%%%%%%%%%%%%%%%%%%%%%%%%%%%%%%%%

\begin{thebibliography}{999}

\bibitem{Epstein}
P.S. Epstein, On the resistance experienced by spheres in their motion through gases, {\em Phys. Rev.} {\bf 23}, pp. 710-733 (1924). 

\bibitem{tsien}
H.S. Tsien, Super-aerodynamics, mechanics of rarefied gases,  {\em J. Aeronaut. Sci.} {\bf 1} (1946). 

\bibitem{ashley} 
H. Ashley, Applications of the theory of free molecule flow to aeronautics, {\em J. Aeronaut. Sci.} {\bf 16}, pp. 95 (1949).  

\bibitem{Heineman}
M. Heineman, Theory of drag in highly rarefied gases, {\em Commun. Pur. Appl. Math.} {\bf 1} (3) (1948).

\bibitem{Stalder}
J.R. Stalder, and V.J. Zurick, Theoretical aerodynamic characteristics of bodies in a free-molecule-flow field, NACA Technical Note 2453, 1951. 

%% cone, sphere, cylinder, plate, no CM, not CQ.  
\bibitem{Gustafson}
W.A. Gustafson, The Newtonian diffuse method for computing aerodynamic force, LMSD T.M. 5132, Lockheed Aircraft Corporation Missile and Space Division, August 28, 1958. 

\bibitem{sentman}
L.H. Sentman, Free molecular flow theory and its application of the determination of aerodynamic forces, Technical Report AD 265409,  LMCS-448514, Lockheed Missiles \& Space Company, a division of Lockheed aircraft corporation, Sunnyvale, California, 1961.

%%Hypersonic Newtonian flow; give up normal direction velocity, and slip through tangently. with cone, cylinder, sphere,  and  Cd, Cf, Cm 
\bibitem{Stalder_Jukoff}
J.R. Stalder,  and D., Jukoff, Heat transfer to bodies traveling at high speed in the upper atmosphere, {\em J. Aeronaut. Sci.} {\bf 15}, (7), 381-391 (1948). 

\bibitem{Oppenheim}
A.K. Oppenheim, Generalized theory of convective heat transfer in a free molecule flow, {\em J. aeronaut. sci.} {\bf 25}, 49-58, (1953).

\bibitem{Schaaf}
S. A. Schaaf; P. L. Chambre, {\em Flow of Rarefied Gases}, Princeton University Press, Princeton, 1961. 

\bibitem{ctwang}
C.T. Wang, Free molecular flow over a rotating sphere, {\em AIAA J.} {\bf 10} (5), 713-714 (1972).

\bibitem{Storch}
J. A. Storch, Aerodynamic disturbances on spacecraft in free-molecular flow, SMC-TR-03-06, Aerospace Report No. TR-2003(3397)-1 (2002).

\bibitem{disc}
J.V. Sengers, Y. L Wang,  B. Kamgar-Parsi, and J.R. Dorfmana,  Kinetic theory of drag on objects in nearly free molecular flow, {\em Physica A: Stat. Mech. Appli.} {\bf 413}, 409-425 (2014).

\bibitem{cai_jsr}
C. Cai, K. Khasawneh, H. Liu, and W. Wei, Collisionless gas flows over a cylindrical or spherical object, {\em J. Spacecr. Rockets} {\bf 46} (6), 1124-1131 (2009).

\bibitem{cai_plate}
C. Cai and K. Khasawneh, K., Collisionless gas flow over a cryogenic flat plate, {\em J. Vacu. Sci. Technol.}  {\bf 27} (4), 601-610 (2009).

\bibitem{DSMC}
G.A. Bird, {\em Molecular Gas Dynamics and Direct Simulation of Gas Flows} (Clarendon Press, Oxford, 1994).

\bibitem{SanjeeviS}
S.K. Sanjeevi, J.A. Kuipers, and J.T. Paddinga, Drag, lift and torque correlations for non-spherical particles from Stokes limit to high Reynolds numbers, {\em Int. J. Multiph. Flow}, {\bf 106},  Pages 325-337 (2018).

\bibitem{Ouchene}
R. Ouchene, M. Khalij, B. Arcen, and Anne Tanière, A new set of correlations of drag, lift and torque coefficients for non-spherical particles and large Reynolds numbers, {\em Powder Technol}. {\bf 303}, 33-43 (2016)

\bibitem{Happel}
J. Happel, H. Brenner, {\em Low Reynolds Number Hydrodynamics}, (Prentice-Hall, Englewood Cliffs, 1965).

\bibitem{Zastawny}
M. Zastawny, G. Mallouppas, F. Zhao, and B. Wachem, Derivation of drag and lift force and torque coefficients for non-spherical particles in flows, {\em Int. J. Multiph. Flow}. {\bf 39}, Pages 227-239 (2012).

\bibitem{Ouchene1}
R. Ouchene, M. Khalij, A. Tanière, and B. Arcen, Drag, lift and torque coefficients for ellipsoidal particles: from low to moderate particle Reynolds numbers, {\em Computers \& Fluids}. {\bf 113}, Pages 53-64 (2015).

\bibitem{Channapan_POF}
A.K. Chinnappan, R. Kumar,  V.K. Arghode, and R.S. Myong, Transport dynamics of an ellipsoidal particle in free molecular gas flow regime, {\em Phys. Fluids}. {\bf 31}, 037104 (2019).

\bibitem{Loth}
E. Loth, Drag of non-spherical solid particles of regular and irregular shape, {\em Powder Technol.} {\bf 182}, 342–353 (2008).

\bibitem{ellipse}
%https://en.wikipedia.org/wiki/Ellipse, last visit on Jan. 20, 2020.
M.H.  Protter, and  C.B. Morrey, {\em College Calculus with Analytic Geometry (2nd ed.)} (Addison-Wesley, Boston, 1970). 

\bibitem{impingement}
C. Cai, X. He, and K. Zhang, Comprehensive studies on rarefied jet and jet impingement flows with gaskinetic methods, {\em Commu. Compu. Phys.} {\bf 23} (3), 712-741 (2017).

\bibitem{Kogan}
M.N. Kogan, {\em Rarefied Gas Dynamics}  (Plenum Press, New York, 1969).

\bibitem{GRASP}
H. Liu, C. Cai, and  C. Zou, An object-oriented serial implementation of a DSMC simulation package,  {\em Comput. \& Fluids}. {\bf 57}, 66-75 (2012).

%%%%%%%%%%%%%%%%%%%%%%%%%%%%%%%%%%%%%%%%%%%%%%%%%%
\bibitem{Mando}
M. Mandø and L. Rosendahl, On the motion of non-spherical particles at high Reynolds number, {\em Powder Technol.} {\bf 202} (1-3), 1–13 (2010).

\bibitem{Rosendahl}
L. Rosendahl, Using a multi-parameter particle shape description to predict the motion of non-spherical particle shapes in swirling flow, {\em Appl. Math. Model.} {\bf 24} (1), 11–25 (2000).

\bibitem{Yin}
C. Yin, L. Rosendahl, S. K. Knudsen, and H. Srensen, Modelling the motion of cylindrical particles in a nonuniform flow, {\em Chem. Eng. Sci.} {\bf 58} (15), 3489–3498 (2003).

\bibitem{Sanjeevi}
S. K. P. Sanjeevi, and J. T. Padding, On the orientational dependence of drag experienced by spheroids, {\em J. Fluid Mech.} (820), R1 (2017). 
%\begin{equation}
%  X_{cp}/L= 0.25 ( 1- \sin^3 \alpha); X_{cp}/L =  0.25 \cos^3 \alpha
%\end{equation}

\bibitem{Taguchi}
S. Taguchi, ``Asymptotic theory of a uniform flow of a rarefied gas past a sphere at low Mach numbers,'' {\em
J. Fluid Mech.} (774), 363-394 (2015). 


\end{thebibliography}
\end{document}